\documentclass[lettersize,journal]{IEEEtran}
\usepackage{amsmath,amsfonts}
\usepackage{array}
\usepackage[caption=false,font=normalsize,labelfont=sf,textfont=sf]{subfig}
\usepackage{textcomp}
\usepackage{stfloats}
\usepackage{url}
\usepackage{verbatim}
\usepackage{graphicx}
\usepackage{cite}
\AtBeginDocument{%
  \providecommand\BibTeX{{%
    \normalfont B\kern-0.5em{\scshape i\kern-0.25em b}\kern-0.8em\TeX}}}
    
\usepackage{array, multirow, multicol}  
\usepackage{enumitem}

\usepackage{tikz}
\usetikzlibrary{shadows,patterns,shapes,arrows,decorations.pathmorphing,backgrounds,positioning,fit,plotmarks,calc,spy,petri,topaths}
\usetikzlibrary{arrows.meta}

\usepackage{pgfplots}
\usepackage{makecell}
\usetikzlibrary{plotmarks} 
\usepackage[ruled, lined, linesnumbered, commentsnumbered, longend]{algorithm2e}
\SetKwFor{ForEach}{foreach}{do}{}
\SetKwIF{If}{ElseIf}{Else}{if}{then}{else if}{else}{}
\usepackage{balance}

\usepgfplotslibrary[groupplots]
\pgfplotsset{compat=1.18,
    /pgfplots/ybar legend/.style={
    /pgfplots/legend image code/.code={%
       \draw[##1,/tikz/.cd,yshift=-0.25em]
        (0cm,0cm) rectangle (4pt,0.6em);},
   },
}
\usetikzlibrary{matrix, patterns, patterns.meta, shapes.geometric}
\tikzdeclarepattern{
  name=mylines,
  parameters={
      \pgfkeysvalueof{/pgf/pattern keys/size},
      \pgfkeysvalueof{/pgf/pattern keys/angle},
      \pgfkeysvalueof{/pgf/pattern keys/line width},
  },
  bounding box={
    (0,-0.5*\pgfkeysvalueof{/pgf/pattern keys/line width}) and
    (\pgfkeysvalueof{/pgf/pattern keys/size},
0.5*\pgfkeysvalueof{/pgf/pattern keys/line width})},
  tile size={(\pgfkeysvalueof{/pgf/pattern keys/size},
\pgfkeysvalueof{/pgf/pattern keys/size})},
  tile transformation={rotate=\pgfkeysvalueof{/pgf/pattern keys/angle}},
  defaults={
    size/.initial=5pt,
    angle/.initial=45,
    line width/.initial=.4pt
  },
  code={
      \draw [line width=\pgfkeysvalueof{/pgf/pattern keys/line width},densely dotted]
        (0,0) -- (\pgfkeysvalueof{/pgf/pattern keys/size},0);
  },
}

\definecolor{blizzardblue}{rgb}{0.67, 0.9, 0.93} 
\definecolor{cadmiumgreen}{rgb}{0.0, 0.42, 0.24} 
\definecolor{lightgray}{rgb}{0.83, 0.83, 0.83} 
\definecolor{deepskyblue}{rgb}{0.0, 0.75, 1.0} 
\definecolor{forestgreen}{rgb}{0.13, 0.55, 0.13} 
\definecolor{lightcyan}{rgb}{0.88, 1.0, 1.0} 
\definecolor{slategray}{rgb}{0.44, 0.5, 0.56} 
\definecolor{limegreen}{rgb}{0.2, 0.8, 0.2} 
\definecolor{mediumblue}{rgb}{0.0, 0.0, 0.8} 
\definecolor{seagreen}{rgb}{0.18, 0.55, 0.34} 
\definecolor{darkgoldenrod}{rgb}{0.72, 0.53, 0.04}
\definecolor{darkmagenta}{rgb}{0.55, 0.0, 0.55}
\definecolor{darkseagreen}{rgb}{0.56, 0.74, 0.56}
\definecolor{cerulean}{rgb}{0.0, 0.48, 0.65}
\definecolor{darkseagreen}{rgb}{0.56, 0.74, 0.56}
\definecolor{darkspringgreen}{rgb}{0.09, 0.45, 0.27}
\definecolor{darkturquoise}{rgb}{0.0, 0.81, 0.82}
\definecolor{denim}{rgb}{0.08, 0.38, 0.74}
\definecolor{dollarbill}{rgb}{0.52, 0.73, 0.4}
\definecolor{darkgreen}{rgb}{0.0, 0.2, 0.13}
\definecolor{darkblue}{rgb}{0.0, 0.0, 0.55}
\definecolor{calpolypomonagreen}{rgb}{0.12, 0.3, 0.17}
\definecolor{darkcyan}{rgb}{0.0, 0.55, 0.55}
\definecolor{ferngreen}{rgb}{0.31, 0.47, 0.26}
\definecolor{forestgreen(traditional)}{rgb}{0.0, 0.27, 0.13}
\definecolor{navyblue}{rgb}{0.0, 0.0, 0.5} 
\definecolor{forestgreen}{rgb}{0.0, 0.27, 0.13} 
\definecolor{burgundy}{rgb}{0.5, 0.0, 0.13} 
\definecolor{slategray}{rgb}{0.44, 0.5, 0.56} 
\definecolor{charcoal}{rgb}{0.21, 0.27, 0.31} 
\definecolor{maroon}{rgb}{0.5, 0.0, 0.0} 
\definecolor{darkteal}{rgb}{0.0, 0.2, 0.2} 
\definecolor{taupe}{rgb}{0.28, 0.24, 0.2} 
\definecolor{olivegreen}{rgb}{0.33, 0.42, 0.18} 
\definecolor{steelblue}{rgb}{0.27, 0.51, 0.71} 
\definecolor{bluebell}{rgb}{0.301, 0.341, 0.690} 
\definecolor{aliceblue}{rgb}{0.953, 0.953, 0.953} 
\definecolor{vermilion}{rgb}{0.988, 0.392, 0.298} 
\definecolor{corduroy}{rgb}{0.600, 0.600, 0.600} 
\definecolor{brickred}{rgb}{0.804, 0.416, 0.518} 
\definecolor{eucalyptus}{RGB}{59, 110, 100}
\definecolor{cadet}{RGB}{66, 123, 114}
\definecolor{spray}{RGB}{141, 192, 179}
\definecolor{portage}{RGB}{123, 129, 164}
\definecolor{palelilac}{RGB}{209, 200, 218}
\definecolor{laurelgreen}{RGB}{137, 138, 124}
\definecolor{satPointsColor}{HTML}{D7191C}
\definecolor{unsatPointsColor}{HTML}{FDAE61}
\definecolor{timedOutPointsColor}{HTML}{ABDDA4}
\definecolor{rendering}{HTML}{2B83BA}
\definecolor{lightBlue}{RGB}{173, 216, 230}   
\definecolor{lightRed}{RGB}{255, 182, 193}    
\definecolor{lightYellow}{RGB}{255, 255, 224} 
\definecolor{lightGreen}{RGB}{144, 238, 144}  
\definecolor{purple}{RGB}{128, 0, 128}        
\definecolor{gray}{RGB}{128, 128, 128}        
\definecolor{deeperBlue}{RGB}{100, 149, 237}  
\definecolor{deeperRed}{RGB}{220, 20, 60}     
\definecolor{deeperYellow}{RGB}{238, 220, 130} 
\definecolor{deeperGreen}{RGB}{34, 139, 34}   
\definecolor{deeperPurple}{RGB}{104, 34, 139} 
\definecolor{deeperGray}{RGB}{105, 105, 105}  
\definecolor{richYellow}{RGB}{204, 204, 0}  
\definecolor{deeperorange}{RGB}{205,85,0}

\definecolor{lightpink}{HTML}{E89DA0}
\definecolor{lightblue}{HTML}{88CEE6}
\definecolor{peach}{HTML}{F6C8A8}
\definecolor{lightgreen}{HTML}{B2D3A4}
\definecolor{olivegreen}{HTML}{9FBA95}
\definecolor{palepink}{HTML}{E6CECF}
\definecolor{lilac}{HTML}{B696B6}
\definecolor{teal}{HTML}{80C1C4}

\usepackage{pifont}
\newcommand{\cmark}{\textcolor{cadmiumgreen}{\ding{51}}}
\newcommand{\xmark}{\textcolor{deeperRed}{\ding{55}}}

\usepackage[colorinlistoftodos]{todonotes}

\begin{document}

\title{Para-Pipe: Exploiting Hierarchical Operator \\ Parallelism of ML Computational Graphs on SoCs}

\author{Yujie Zhang, Huiying Lan, Ehsan Aghapour, Zhiyuan Ning,\\ Peng Zan, Weidong Shao, Anuj Pathania, and Tulika Mitra
\thanks{This work was supported by the National Research Foundation, Singapore under its Competitive Research Program Award NRF-CRP23-2019-0003 and by a gift from Black Sesame Technologies. This article was recommended by Associate Editor A. Kumar. (\textit{Corresponding author: Tulika Mitra.})}

\thanks{\copyright~2025 IEEE. Personal use is permitted, but republication/redistribution requires IEEE permission.}%
\thanks{\textbf{This article was published in the IEEE Transactions on Computer-Aided Design of Integrated Circuits and Systems (TCAD), vol.~44, no.~12, pp.~4472--4485, Dec.~2025, doi: 10.1109/TCAD.2025.3568348.}}%

\thanks{Yujie Zhang, Huiying Lan, and Tulika Mitra are
with the School of Computing, National University of Singapore, Singapore (e-mail: zyujie@comp.nus.edu.sg; hy.lan@nus.edu.sg; tulika@comp.nus.edu.sg). 

Ehsan Aghapour and Anuj Pathania are with Parallel Computer Systems, the University of Amsterdam, 1098 XH Amsterdam, The Netherlands (e-mail: e.aghapour@uva.nl; a.pathania@uva.nl).

Zhiyuan Ning, Peng Zan, and Weidong Shao are with the Information Technology Department, Black Sesame Technologies, San Jose, CA 95131, USA (e-mail: thomas.ning@bst.ai; peng.zan@bst.ai; david.shao@bst.ai).}
}



\maketitle
\thispagestyle{empty}

\begin{abstract}
As edge-based deep learning applications become more complex, optimizing performance on heterogeneous System-on-Chips (SoCs) presents unique challenges. 
Traditional pipelining techniques distributing the computation across different on-chip processing units, while effective for throughput, do not address the latency demands posed by modern neural networks with complex interdependencies and extensive operator parallelism. There is a potential in leveraging operator parallelism to enable concurrent execution across multiple processing units, thereby reducing inference latency. However, prioritizing pipelining or parallel execution often necessitates a compromise, where optimizing one performance metric adversely impacts the other.

This paper introduces {\em Para-Pipe}, a hierarchical mapping framework that integrates intra- and inter-stage operator parallelism within a pipelined architecture.
{\em Para-Pipe} navigates the trade-off between throughput and latency by selectively fine-tuning parallelism levels within and across pipeline stages. 
This strategy can significantly reduce inter-processor communication overhead, significantly improving energy efficiency.
Our evaluation demonstrates that {\em Para-Pipe} generates multiple Pareto-optimal configurations, achieving a balance between throughput and latency on an Amlogic SoC equipped with \textit{ARM big.LITTLE} CPUs and GPU, as well as the Black Sesame Technology SoC featuring a deep learning accelerator and two DSPs. 
More importantly, throughput-optimized configurations under {\em Para-Pipe} on Amlogic SoC show an average energy efficiency improvement of 11.0\% over purely pipelined strategies and 23.3\% relative to non-pipelined parallel execution. 
\end{abstract}


\section{Introduction}
\IEEEPARstart{E}{dge} devices increasingly utilize heterogeneous multiprocessor System-on-Chips (SoCs), which comprise a variety of computing units, including CPUs, GPUs, and specialized hardware accelerators, each distinguished by unique power-performance characteristics~\cite{mitra2015heterogeneous}. Despite the capabilities of these diverse units, mainstream machine learning (ML) frameworks such as PyTorch~\cite{paszke2019pytorch}, TVM~\cite{chen2018tvm}, and the \textit{ARM Compute Library (ARM-CL)}~\cite{ARM-CL} frequently underutilize them. Typically, these frameworks leverage the highest-performance or most energy-efficient unit for sequential task execution, simplifying software complexity. While this approach facilitates programming and deployment across different neural networks on edge devices, it neglects the advantages of simultaneously utilizing multiple compute units, which could significantly improve execution speed and energy efficiency.

\begin{figure}[t]
	\raggedleft
    \includegraphics[width=0.5\textwidth, 
    height=0.15\textwidth
    ]{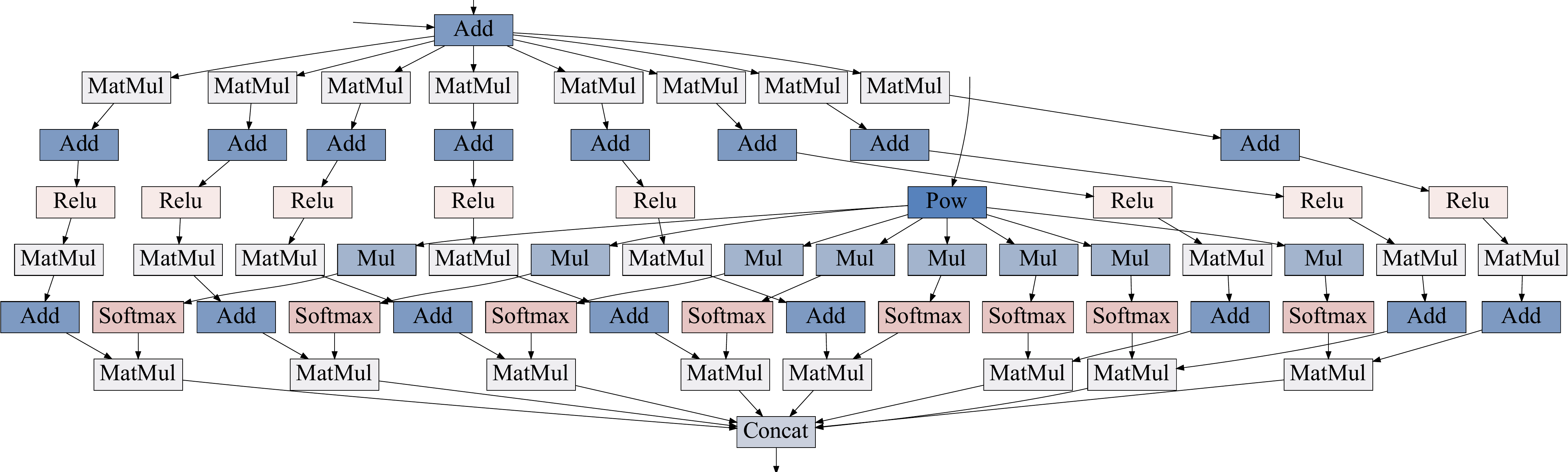}
    \caption{A portion of the PETR network architecture to illustrate intensive operator parallelism~\cite{liu2022petr}.}
    \label{fig:model-structure-demo}
\end{figure}

To better harness diverse resources, pipelining techniques partition ML computational graphs and allocate segments to specific compute units on the SoC, effectively maximizing throughput by fully utilizing on-chip resources for concurrent processing of streaming inputs~\cite{wang2019high,wu2020pipeline,minakova2020combining,aghapour2024arm,wang2024flexi,aghapour2023pipelined}.
These methods are primarily designed for traditional convolutional neural networks like MobileNet. Their predominantly sequential architecture is conducive to straightforward partitioning. However, the growing complexity and inter-dependencies among operators in modern neural networks present substantial challenges for effective pipelining. Consider, for example, PETR network~\cite{liu2022petr}, a spatiotemporal transformer-based encoder for autonomous driving perception tasks, with partial architecture illustrated in Fig.~\ref{fig:model-structure-demo}. Identifying optimal partition points within such complex networks is increasingly challenging due to the expanded number of operations and potential partition points. Moreover, this approach can lead to increased latency per frame due to enhanced inter-pipeline stage communication overhead and suboptimal use of concurrent operator execution within individual frames. Such latency increases are particularly detrimental in real-time decision-making applications, including autonomous driving~\cite{li2022bevformer} and some IoT devices~\cite{xiao2018iot,zantalis2019review}, where prompt response times are imperative.

Despite their complexity, modern neural networks are ripe with opportunities for concurrent operator execution to improve latency markedly. 
For instance, the InceptionNet family~\cite{szegedy2015going} employs dimension-reduced inception modules that process the same input with filters of varying scales in parallel. Similarly, transformer-based networks utilize multiple attention heads within self-attention mechanisms to parallel-process the same input sequence. 
These computational designs demonstrate considerable inter-operator parallelism, devoid of complex data dependencies, suggesting that strategic partitioning and allocation of operations to multiple processing units for truly parallel execution could substantially enhance latency. However, prioritizing latency in single-frame acceleration by engaging all available units~\cite{zhang2021duet,tumeo2008ant,bender1996milp,topcuoglu2002performance,zhang2015maximizing} undermines the throughput benefits of pipelining.

Achieving a balanced trade-off between latency and throughput is critical for optimizing performance in edge-based deep learning applications. 
Enhancements in throughput reduce overall execution times for processing large-scale inputs, while latency improvements decrease delays for individual frames. In applications like video surveillance~\cite{tsakanikas2018video,cheng2021vitrack,chen2019distributed}, where both rapid processing of extensive data and swift handling of single frames are essential, optimizing both throughput and latency concurrently is imperative. In these instances, standard pipelining or parallel execution strategies that promote throughput at the expense of latency—or vice versa—are suboptimal.

To tailor and optimize the balance between latency and throughput in neural networks characterized by intensive operator parallelism on heterogeneous SoCs, we introduce {\em Para-Pipe}. This hybrid framework integrates pipelining with parallel execution through a hierarchical structure. 
{\em Para-Pipe} systematically identifies neural network subgraphs suitable for parallel processing (e.g., GoogLeNet inception modules) and allocates them across multiple units to maximize this parallelism. For subgraphs exhibiting a lower operator parallelism, it assigns them to individual units for sequential processing. 
This dual approach effectively separates sequential and parallel processing into distinct pipeline stages, thereby facilitating hierarchical exploitation of inter-operator parallelism within and across stages. {\em Para-Pipe} can selectively modulate intra- and inter-stage operator parallelism levels, finely tuning the latency and throughput balance. Correspondingly, it regulates the cost associated with inter-processor communication within and among stages, optimizing for significant reductions and markedly enhancing energy efficiency—a critical issue in embedded heterogeneous computing environments, especially in multiprocessor SoCs~\cite{zhang2015maximizing,bouzidi2023map,chen2024ped}. Para-Pipe stands out in terms of two aspects:

\textit{Para-Pipe optimizes the latency-throughput trade-off.}
{\em Para-Pipe} provides the flexibility to adjust the workload of parallel operator execution, enabling tailored optimization to suit specific computational requirements.
Generally, latency decreases when fewer pipeline stages and more parallel computing resources are allocated to each stage, while throughput benefits from a diminished workload per stage by adding more pipeline stages.
The lowest latency is achieved by engaging all available processors for single-frame inference without pipelining. Meanwhile, the highest throughput is achieved for multi-frame inference by creating multiple pipeline stages so that the execution time of the longest stage is at a minimum. 
Crucially,  based on the workload customization of parallel operator execution within stages, \textit{Para-Pipe} effectively manages the trade-off between latency and throughput.

\textit{Para-Pipe enhances energy efficiency in heterogeneous SoCs.}
Relative to parallelism-only methods, {\em Para-Pipe}'s hybrid execution mode significantly reduces inter-processor communication costs by allocating specific subgraphs to stages managed by a single processor for sequential execution. Moreover, compared to the pipeline-only setup, {\em Para-Pipe} diminishes the necessity for inter-stage communication and coordination by streamlining the number of pipeline stages. Nonetheless, the parallel execution stages are not entirely exempt from incurring inter-processor communication costs. Through the judicious selection of processors for concurrent execution within a single stage, \textit{Para-Pipe} effectively mitigates such overhead, thereby augmenting energy efficiency.

We evaluate our framework across four InceptionNet family and two transformer-based models on two real SoCs, an Amlogic A311D SoC~\cite{AMLOGIC_A311D}, equipped with dual CPU clusters and a GPU, in addition to the Black Sesame Technology (BST) SoC including an NPU and two DSPs. The results reveal that {\em Para-Pipe} generates Pareto-optimal mapping options tailored to various throughput and latency priorities. Notably, hybrid configurations optimized for throughput on Amlogic SoC achieve an average energy efficiency improvement of 11.0\% compared to purely pipelined strategies and 23.3\% relative to non-pipelined parallel execution.

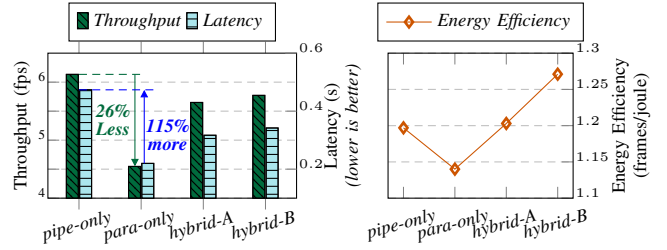
\begin{figure}[t]
\raggedleft
    \begin{tikzpicture}
\centering
    \pgfplotsset{
        lat/.style={
            fill=blizzardblue,
            postaction={pattern=horizontal lines},
        },
    }
    \begin{groupplot}[
        group style={
            group name=my plots A,
            group size=1 by 1,
            xlabels at=edge bottom,
            xticklabels at=edge bottom,
            y descriptions at=edge left,
            ylabels at=edge left,
            yticklabels at=edge left,
        },
        xtick=data,
        x tick label style={font=\scriptsize,rotate=15,text height=0pt,text width=1.5cm,align=center,inner sep=1pt,outer sep=1pt},
        xlabel style={font=\scriptsize, align=center, text height=0pt,inner sep=2pt,outer sep=2pt},
        xticklabels from table={data/perf_diff_inv3.dat}{Model},
        ybar, /pgf/bar shift=-2.5pt, 
        ymin=4, ymax=6.5,
        ytick pos=left,
        y tick label style={font=\tiny,rotate=0,text height=1pt,inner sep=1pt,outer sep=0pt},
        ylabel style={font=\scriptsize, align=center, text height=0pt,inner sep=0pt,outer sep=2pt},]

        \nextgroupplot[width=0.55\columnwidth,, height=3.5cm,
                        ylabel={Throughput (fps)},
                        ymajorgrids=true,
                        grid style=densely dashed,
                        bar width=4.5pt,   
                        enlarge x limits={abs=0.5},
                        ytick={4,4.5,5,5.5,6},
                        yticklabels={4,,5,,6},
                        typeset ticklabels with strut,
                        legend columns=2,
                        legend style={at={(0.5,1.08,1)}, anchor=south, font=\scriptsize},]
        \addplot[fill=cadmiumgreen,postaction={pattern=north west lines}]
                table[x expr = \coordindex, y = Throughput,]{data/perf_diff_inv3.dat}; 
        \addlegendentry{\em Throughput}
        \addlegendimage{lat,ybar legend}
        \addlegendentry{\em Latency}

        \draw[cadmiumgreen, densely dashed] (0,6.134) -- (1,6.134);
        \draw (0.55,5.5) node[align=center] {\scriptsize \bfseries \textcolor{cadmiumgreen}{{\em 26\%}}};
        \draw (0.55,5.2) node[align=center] {\scriptsize \bfseries \textcolor{cadmiumgreen}{{\em Less}}};
        \draw[-{Triangle[fill=cadmiumgreen,length=1mm,width=1mm]}, cadmiumgreen,] (0.9,6.134) -- (0.9,4.547);

    \end{groupplot}  

    \begin{groupplot}[
        group style={
            group name=my plots A,
            group size=1 by 1,
            y descriptions at=edge right,
            ylabels at=edge right,
            yticklabels at=edge right,
        },
        xtick=\empty, axis line style=transparent,
        ytick pos=right,
        y tick label style={font=\tiny,rotate=0,text height=1pt,inner sep=1pt,outer sep=0pt},
        ylabel style={font=\scriptsize, align=center, text height=0pt,inner sep=10pt,outer sep=2pt},
        ylabel={Latency (s)\\\textit{(lower is better)}},
        ybar, /pgf/bar shift=2.5pt, 
        ymin=0.1, ymax=0.6, ]
        
        \nextgroupplot[width=0.55\columnwidth,, height=3.5cm,
                        bar width=4.5pt,
                        enlarge x limits={abs=0.5},
                        typeset ticklabels with strut,]
        \addplot[fill=blizzardblue,postaction={pattern=horizontal lines}]
                table[x expr = \coordindex, y = Latency,]{data/perf_diff_inv3.dat};

        \draw[blue, densely dashed] (0,0.4736) -- (1.2,0.4736);
        \draw (1.43,0.35) node[align=center] {\scriptsize \bfseries \textcolor{blue}{{\em 115\%}}};
        \draw (1.43,0.28) node[align=center] {\scriptsize \bfseries \textcolor{blue}{{\em more}}};
        \draw[{Triangle[fill=blue,length=1mm,width=1mm]}-, blue,] (1.05,0.4736) -- (1.05,0.2199);

    \end{groupplot}  

    \begin{scope}[xshift=4.5cm]
        \begin{groupplot}[
            group style={
                group name=my plots B,
                group size=1 by 1,
                xlabels at=edge bottom,
                xticklabels at=edge bottom,
                y descriptions at=edge right,
                ylabels at=edge right,
                yticklabels at=edge right,
            },
            xtick=data,
            x tick label style={font=\scriptsize,rotate=15,text height=0pt,text width=1.5cm,align=center,inner sep=4pt,outer sep=1pt},
            xlabel style={font=\scriptsize, align=center, text height=0pt,inner sep=2pt,outer sep=2pt},
            xticklabels from table={data/perf_diff_inv3.dat}{Model},
            y tick label style={font=\tiny,rotate=0,text height=1pt,inner sep=1pt,outer sep=0pt},
            ytick pos=right,
            ylabel style={font=\scriptsize, align=center, text height=0pt,inner sep=11pt,outer sep=2pt},]
    
            \nextgroupplot[width=0.455\columnwidth,, height=3.5cm,
                            ymajorgrids=true,
                            grid style=densely dashed, 
                            ymin=1.1, ymax=1.3,
                            ylabel={Energy Efficiency\\(frames/joule)},             
                            typeset ticklabels with strut,
                            legend columns=1,
                            legend style={at={(0.5,1.08,1)}, anchor=south, font=\scriptsize},]
                \addplot+[mark=diamond, deeperorange, mark size=2pt, mark options={line width=1pt}]table[x expr = \coordindex,y = Energyefficiency,]{data/perf_diff_inv3.dat};
                \addlegendentry{\em Energy Efficiency}
                
        \end{groupplot}  
    \end{scope}
\end{tikzpicture}
    \vspace*{-2mm}
    \caption{Latency-throughput trade-offs and energy efficiency for scheduling and mapping options for Inception-v3~\cite{szegedy2016rethinking} on Amlogic SoC~\cite{AMLOGIC_A311D}. hybrid-A and hybrid-B combine pipelined and parallel execution with varying degrees of parallel operator execution.}
    \label{fig:latency-throughput-trade-off-v3}
\end{figure}

\begin{figure*}[t!]
    \centering
    \includegraphics[width=\textwidth]{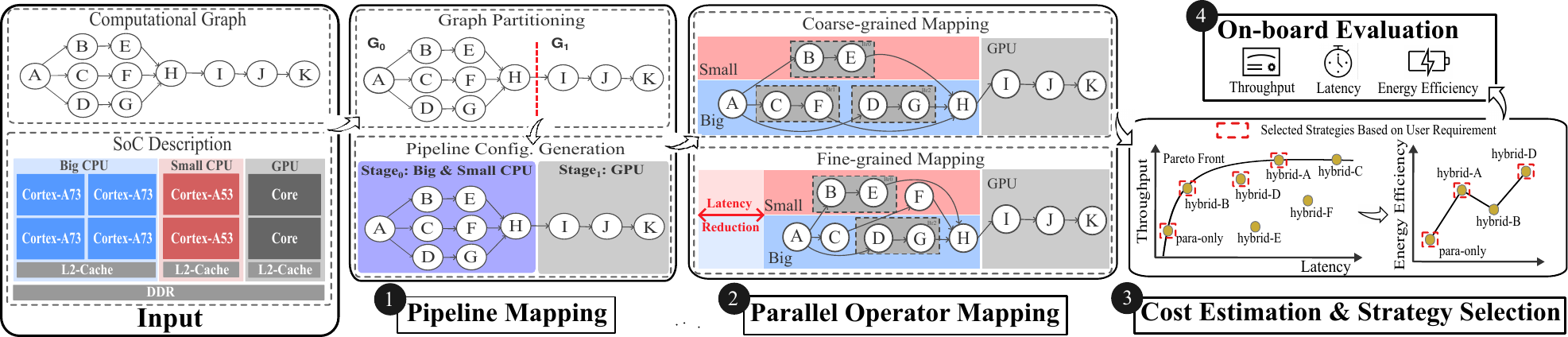}
    \caption{{\em Para-Pipe} framework: High-level overview.}
    \label{fig:Para-Pipe}
\end{figure*} 

\textbf{Motivating Example:}
Fig.~\ref{fig:latency-throughput-trade-off-v3} shows performance trade-offs associated with various mapping options during Inception-v3 network~\cite{szegedy2016rethinking} inference on an Amlogic SoC. 
Methods focusing solely on pipelining (`pipe-only') or parallel execution (`para-only') show significant gains in throughput and latency, respectively. However, these approaches also reveal notable degradation in other metrics, a 115\% increase in latency for `pipe-only' and a 26\% decrease in throughput for `para-only'.
In contrast, the `hybrid-A' and `hybrid-B' strategies, which integrate pipelining with parallel execution, offer a more balanced performance by modulating the degree of inter-operator parallelism within the pipeline stages.
Notably, the `hybrid-A' configuration leads to a 7.9\% reduction in throughput but concurrently secures a substantial latency improvement of 33.1\%, compared to the `pipe-only' approach. 
Furthermore, the hybrid methods enhance the energy efficiency of the SoC; for instance, the `hybrid-B' achieves improvements of 6.2\% and 11.5\% over the `pipe-only' approach and the `para-only' approach, respectively.

Our key contributions are summarized as follows: 
\begin{itemize}[noitemsep,topsep=0pt]
    \item We introduce {\em Para-Pipe}, a partitioning and scheduling framework designed to optimize the inference of modern ML models on heterogeneous SoCs. {\em Para-Pipe} adeptly balances the trade-offs between latency and throughput while significantly enhancing energy efficiency.
    \item We propose a two-stage hierarchical mapping approach that effectively leverages operator parallelism within a pipelined structure, facilitating the development of robust strategies for large-scale models.
    \item We develop two Integer Linear Programming (ILP)-based parallel operator mapping algorithms with distinct granularities, efficiently balancing mapping precision and mapping time. 
    \item We implement {\em Para-Pipe} runtime with \textit{ARM Compute Library} and evaluate its performance on Amlogic and BST SoC to demonstrate versatility and applicability on real platforms.  
\end{itemize}

\section{Overview}

{\em Para-Pipe} features two levels of mapping: pipeline-level (Section~\ref{sec:pipe-mapping}) and operator-level (Section~\ref{sec:para-mapping}) as illustrated in Fig.~\ref{fig:Para-Pipe}.  
In contrast to traditional DAG (Directed Acyclic Graph) mapping algorithms~\cite{tumeo2008ant,bender1996milp,topcuoglu2002performance}, which are designed for generic task graphs and conventional pipelining techniques~\cite{wang2019high,wu2020pipeline,minakova2020combining,aghapour2024arm,wang2024flexi,aghapour2023pipelined}, our approach exploits the well-defined structures of most ML computational graphs. This enables us to establish clear boundaries for pipeline stages, optimizing the mapping process more effectively.
Given a computational graph sourced from ML frameworks like PyTorch and TensorFlow, along with an SoC specification, Pipeline Mapper identifies subgraphs that exhibit dense operator parallelism and aligns them with a subset of compute engines to form effective pipeline stages. This approach simplifies the graph structure, making it more linear and accessible to partition into pipeline stages, enabling both intra-stage operator and inter-stage pipeline parallelism.

Operator Mapper, tailored for each pipeline stage, assigns operations to compute engines to maximize parallelism and reduce latency. It minimizes costs via ILP. Two algorithms with different granularities are introduced: a coarse-grained algorithm treating a branch as the mapping unit and a fine-grained approach mapping each operator individually.
This dual-granularity mapping strategy ensures both generality and efficiency, leveraging multi-branch structures to expedite ILP problem resolution. 
A cost estimator (Section~\ref{sec:cost_selection}) is included to model the computational and communication costs required for the ILP formulation.

{\em Para-Pipe} evaluates all available strategies using the cost estimator to determine network inference latency, throughput, and energy efficiency. It selects mapping strategies exhibiting Pareto optimality (Section~\ref{sec:strategy_select}), offering users multiple choices aligned with their application requirements. 
Currently, {\em Para-Pipe} operates as a one-time static process, with no runtime dynamicity incorporated.
The performance of these mappings is subsequently validated on the real SoC platform using the {\em Para-Pipe} runtime (Section~\ref{sec:amlogic_runtime}).
\section{Pipeline Mapping}
\label{sec:pipe-mapping}
\subsection{Graph Partitioning}
\label{sec:graph-part}

\begin{algorithm}[t]
\small
\caption{\textbf{Graph Partitioning}}
\label{alg:graph-partitioner}
\KwIn{Computational graph $G$, graph inputs $inputs$, graph outputs $outputs$.}
\KwOut{Subgraph list $subgraphs$.}

$subgraphs \gets \emptyset$; $sinks \gets outputs$\;

$G \gets$ \textbf{\em mergeNodes}$(G)$\; \label{algline:merge_nodes}
\If{\textbf{isLinear}$(G)$}{ \label{algline:check_if_linear_s}
    \ForEach{$node_{merged} \in G$}{
        $G_{sub} \gets \textbf{\textit{createSubgraph}}(node_{merged})$\;
        $subgraphs.\text{add}(G_{sub})$\;
    }
    \Return{$subgraphs$}\;
}\label{algline:check_if_linear_d}

\While{$sinks \neq \emptyset$}{
    $sink \gets sinks.\text{pop()}$\;
    $N_{go} \gets \{sink\}$; $N_{stop}, sink_{next}, G_{sub} \gets \emptyset$\;

    \While{$sink_{next} = \emptyset$ \textbf{and} $N_{go} \neq \emptyset$}{\label{algline:tracing_0}
        $B_{end} \gets \emptyset$\;
        \ForEach{$node_{go} \in N_{go}$}{
            $B_{end} \gets$ \textbf{\em traceToBranchEnd}$(node_{go}, G_{sub})$\;\label{algline:tracebranchends}
        }
        $B_{end} \gets B_{end} \cup N_{stop}$\;
        $sink_{next} \gets$ \textbf{\em isTerminalNode}$(B_{end}, inputs)$\;\label{algline:isSameParent}
        $N_{stop}, N_{go} \gets$ \textbf{\em isExtendable}$(B_{end})$\;\label{algline:isExtensible}
    }
    \eIf{$sink_{next} = \emptyset$}{
        $sinks.\text{add}(sink)$\;\label{algline:traceagain}
    }{
        $sinks.\text{add}(sink_{next})$;
        $subgraphs.\text{add}(G_{sub})$\;\label{algline:subgraph-gen}
    }
}
\end{algorithm}
Modern neural networks often exhibit irregular structures characterized by extensive operator volumes and complex interconnections. To manage this complexity, we partition the computational graph $G$ into non-overlapping, topologically ordered subgraphs, denoted as $G_{1}, G_{2}, ...$.
Each subgraph exhibits either a linear chain of operator nodes or a fan-in structure. A fan-in structure is characterized by the presence of a single sink node to which all source nodes within the subgraph ultimately connect. These source nodes are either graph inputs or share a common parent node, which serves as the sink node for the preceding subgraph. Fan-in structures are prevalent in contemporary neural networks, such as the InceptionNet family and transformer-based models, where they facilitate processing image features and sentence structures by consolidating common inputs into a unified output.

Algorithm~\ref{alg:graph-partitioner} elaborates the graph partitioning approach. 
To simplify the computational graph, operator nodes are initially merged wherever feasible to reduce structural complexity. For purely linear networks, the algorithm partitions the graph by treating each operator node as an independent subgraph, thereby broadening the framework's applicability \textit{(Lines~\ref{algline:check_if_linear_s}-\ref{algline:check_if_linear_d})}. For graphs with irregular structures, the algorithm commences by traversing from graph outputs to discern subgraphs adhering to fan-in structures. 
The graph outputs serve as initial sink nodes, recorded in the set $sinks$. The entire graph is iteratively explored until $sinks$ is empty.

To identify subgraphs, two sets, $N_{go}$ and $N_{stop}$, are initialized to track extensible and non-extensible nodes, respectively. $G_{sub}$ maintains nodes traced in the current subgraph. Initially, only the sink node is placed in $N_{go}$, and tracing operations are executed on all nodes within $N_{go}$. 

During tracing, the furthest nodes in the same linear chain as nodes in $N_{go}$ are identified and stored in $B_{end}$ \textit{(Line~\ref{algline:tracebranchends})}. These nodes serve as potential sources. Subsequently, $B_{end}$ is updated with nodes from $N_{stop}$. If nodes in $B_{end}$ represent inputs or share a common parent \textit{(Line~\ref{algline:isSameParent})}, the algorithm designates the parent as the sink node for the next subgraph and generates the subgraph $G_{sub}$ \textit{(Line~\ref{algline:subgraph-gen})}. Otherwise, $N_{go}$ and $N_{stop}$ are updated based on the extensibility of nodes in $B_{end}$: for extensible nodes, this update occurs after all their child nodes have been visited, while for non-extensible nodes, the update happens when at least one child node remains unvisited \textit{(Line~\ref{algline:isExtensible})}.
Tracing persists if extensible nodes exist until termination. If no extensible nodes are found, and the traced nodes cannot form a fan-in structure, the sink node is appended to $sinks$ for later consideration \textit{(Line~\ref{algline:traceagain})}. Ultimately, acquired subgraphs are ordered based on their topological sequence.

\subsection{Stage Configuration Generation}
Consider an SoC with $m$ compute engines, \textit{Para-Pipe} offers pipeline configurations ranging from $1$ (para-only) to $m$ (pipe-only) stages.  
It enumerates processor combinations for each configuration, matching the number of stages while preventing processor overlap and ensuring optimal utilization. Specifically, in each configuration, processors are divided into distinct sets corresponding to each stage. This division guarantees that each processor is uniquely assigned to one set, ensuring a non-overlapping and exclusive allocation of processors across different stages.

Once pipeline stages are linked with specific processor combinations, \textit{Para-Pipe} enumerates potential partition points to generate multiple candidate pipeline configurations, determining the set of sequential subgraphs for each stage. The limited number of on-chip processors and the efficient process of identifying subgraphs ensure minimal time overhead for this enumeration. Notably, this enumeration approach can produce a pipe-only configuration that achieves nearly maximal throughput. 
Here, a stage is denoted as $(G_{l,k}, P)$, where $l \le k$, with $G_{l,k}$ representing a graph composed of subgraphs $G_{l}, G_{l+1}, ..., G_{k}$ executed on the processor set $P$. In the case of the Amlogic SoC example in Fig.~\ref{fig:Para-Pipe}, the initial stage employs two CPU clusters for operator parallelism within the first subgraph, followed by the processing of the second subgraph by a GPU in the subsequent stage.
A Parallel Operator Mapping pass is subsequently applied to devise a parallel mapping strategy for the first subgraph.

\section{Parallel Operator Mapping}
\label{sec:para-mapping}
We introduce two ILP-based algorithms operating at distinct granularity. The coarse-grained method, suitable for graphs with independent branch structures, provides solutions less susceptible to execution jitters and synchronization overhead very fast. Conversely, the fine-grained approach, unconstrained by graph structure, yields superior mapping results but with longer solving times. 
For example, as depicted in Fig.~\ref{fig:Para-Pipe}, the fine-grained mapping approach, in contrast to the coarse-grained method, allows for the segmentation of branches and the distribution of operators across CPU clusters, effectively minimizing processor idle time and thus reducing latency.

\subsection{Coarse-grained Mapping}
This mapper allocates operators within a single branch to the same processor to minimize potential communication overhead. Given a stage configuration $(G_{l,k},P)$, we employ it to generate mapping strategies for subgraphs $G_{l}, G_{l+1}, ..., G_{k}$ sequentially. Our objective for each subgraph is to minimize the execution time of the processor with the longest duration.

We begin by extracting independent branches from the subgraph into set $S$. For each branch $s$, we create a binary variable $x_{s}$ in Equation~\eqref{equ:x} to represent its placement strategy. This variable is constrained by $\sum_{p \in P}x_{sp} = 1$, ensuring that a branch is assigned to only one processor. This constraint allows for multiple branches to be allocated to one processor, optimizing the utilization of high-performance processors.
\begin{equation}
    \label{equ:x}
    \begin{array}{l}
    x_{sp} = \begin{cases}
            1, & \text{\parbox[c]{5.3cm}{if computationally demanding path of branch $s$ is assigned to processor $p$}},\\
            0, & \text{otherwise}.
        \end{cases} 
    \end{array}
\end{equation}
In each branch $s$, we quantify the execution time of the computationally intensive path on processor $p$ as $cpc_{sp}$, representing the cumulative sum of execution times for all included operators. The sole sink node in the subgraph is allocated to the highest-performance processor $q$. 
The communication cost between $s$ and the sink node $t$ is denoted as $cmc_{sp} = cmc_{(s_{l},t),pq}$, where $s_{l}$ represents the tail node of the computationally intensive path in branch $s$. The execution cost of $s$ on processor $p$ can be represented as $exec_{sp}=cpc_{sp}+cmc_{sp}$.

The longest execution time duration $L$ is formulated as Equation~\eqref{equ:latency}. To minimize the value of $L$, we utilize ILP techniques to determine optimal values of binary variables $x$, which generates an effective computation path placement for all branches. 
\begin{equation}
    \label{equ:latency}
    L = \max_{p \in P} \sum_{s \in S} exec_{sp}\times x_{sp}
\end{equation}

Our algorithm employs idle or unused processors to address the allocation of operators in nested branches that are not part of the computationally demanding path. It replicates the mapping process to effectively assign these nodes within the nested structures to the appropriate processors, ensuring optimal resource utilization.

\subsection{Fine-grained Mapping}
\label{subsec:ilp}
We first introduce the objective function and the execution time modeling, including computational and communication costs. Then, we describe the constraints to formulate the parallelism limited by the availability of processors. In a graph, the objective function is to minimize the maximum execution time of all last operators. An operator $v$ has $P$ possible mapping processors, which is formulated as a vector of length $P$, each element is a binary variable denoting if $v$ is assigned to processor $p$, described in Equation~\eqref{eq:varx}. $x_{v}$ is constrained by $\sum_{p=1}^{P}x_{vp}  = 1$, which requires an operator to be assigned to only one processor.
\begin{equation}
\label{eq:varx}
  x_{vp} =
  \begin{cases}
  1, & \text{if operator $v$ is assigned to processor $p$}, \\
  0, & \text{otherwise}.
  \end{cases}
\end{equation}

The execution time of $v$ is formulated as Equation~\eqref{eq:exetime}. $st_{v}$ refers to the start time of operator $v$. Operator $v$ starts executing after all predecessors are finished. $exec\_cpc_{v}$ refers to the actual time spent on computation, which is computed as $\sum_{p=1}^{P} x_{vp} \times cpc_{vp}$, and $exec\_{cmc}_{(v,u)}$ is the communication cost for edge $(v,u)$ calculated by $\sum_{p=1}^{P} \sum_{q=1}^{P} y_{(v,u),pq} \times cmc_{(v,u),pq}$.  $cpc$ and $cmc$ are costs profiled offline and estimated with our cost model described in Section~\ref{sec:cost_selection}.
\begin{equation}
\label{eq:exetime}
st_{v} \ge st_{pred} + exec\_cpc_{pred} + exec\_cmc_{(pred,v)} \\
\end{equation}

$y_{(v,u),pq}$ are binary variables representing whether a pair of processors is selected, as shown in Equation~\eqref{eq:y}. They are under similar constraints to $x$ that force only one pair of processors to be selected, $\sum_{p=1}^{P}\sum_{q=1}^{P}y_{(v,u),pq}  = 1$. By introducing this variable, we can formulate the communication cost of the edge $(v, u)$ with the same approach of $exec\_{cpc}$. By multiplying the binary variable and the corresponding cost and summing them together, both the computational and communication time of the graph are formulated.
\begin{equation}
\label{eq:y}
  y_{(v,u),pq} =
  \begin{cases}
  1, & \text{if $x_{vp} \land x_{uq}=1$},\\
  0, & \text{otherwise}.
  \end{cases}
\end{equation}  

Additionally, we formulate the availability of processors to limit the number of operators executing in parallel. This constraint requires that the execution time of two parallel operators assigned to the same processor cannot overlap. For each pair of parallel operators $v$ and $u$, we formulate the constraints in~\eqref{eq:para}, where $ft$ represents the operator finish time.
\begin{equation}
  \label{eq:para}
ft_{v} \le st_{u} \ OR \ ft_{u} \le st_{v}, \text{if $x_{v} = x_{u}$}
\end{equation}

We use the big M technique to express the condition and transform~\eqref{eq:para} to~\eqref{eq:para_expand}.
\begin{equation}
\label{eq:para_expand}
\begin{split}
& ft_{v} - st_{u} \le M \times (2 - x_{vp}-x_{up}) + M \times z_{uvp} \\
& ft_{u} - st_{v} \le M \times (2 - x_{vp}-x_{up}) + M \times (1-z_{uvp}) \\
\end{split}
\end{equation}

$z_{uvp}$ is a set of binary variables introduced to formulate the exclusive condition between the two formulas. If $z_{uvp}=1$, the formula above holds true; otherwise, the below holds true. And $ft_{v}$ is computed with $st_{v}+cpc_{vp}$. We solve this ILP problem with gurobi~\cite{gurobi2022gurobi}.

\section{Cost Estimation \& Strategy Selection}
After the Pipeline Mapping Pass—which details all potential pipeline configurations—and the Parallel Operator Mapping Pass—which establishes mapping strategies within these stages—a comprehensive performance and energy estimation is conducted. {\em Para-Pipe} identifies Pareto-optimal mapping options based on this estimation. Users can then select from these options to find those that best meet their latency, throughput, and energy efficiency preferences.

\subsection{Cost Estimation}
\label{sec:cost_selection}

\subsubsection{Performance}
Within a specific mapping option, model latency is defined as the cumulative execution time of all stages, while throughput is defined as the reciprocal of the execution time of the longest stage. 
For the concurrent execution of processors within a stage, we simulate both the co-execution of processors and the associated communication costs.
On Amlogic SoC, operator computation times are systematically profiled offline.
The costs associated with operator communication are determined based on the time required for data transition and conversion. This includes profiling SDRAM access times for CPUs and GPU, as well as the overhead associated with data conversion tasks like address mapping. These measurements are carried out using benchmarking tools such as TinyMemBench~\cite{siamashka2019tinymembench} and clpeak~\cite{krishnaraj2015clpeak}, which provide precise insights into memory access and data handling efficiencies under various operational scenarios.
For BST SoC, computation and communication costs are derived using its operator simulator, as detailed in Section~\ref{subsec:bst}.

\subsubsection{Energy Efficiency}
Given one mapping strategy, energy efficiency, quantified as inference requests per unit of energy ($frames/joule$), is computed by dividing its throughput by total platform active power. 
Due to the absence of an energy estimation model provided by BST~\cite{BST_Company}, this study evaluates energy efficiency exclusively on the Amlogic SoC. The total active power is determined as the sum of the estimated power consumption of all processors.
It is important to note that memory power is inherently included in the processor power consumption measurements and is therefore not accounted for separately.

We use the model proposed in \cite{pathania2015power} to estimate processor power consumption, expressed in Equation~\eqref{equ:freq_power}. 
At the voltage/frequency level of $V_f/f$, the processor power is the sum of dynamic power $\alpha V_{f}^2 U_{f} f$, where $\alpha$ is capacitance, and $U_f$ is processor utilization, and static power $\beta V_{f}$ with the leakage current $\beta$. 
$\alpha$ and $\beta$ are platform-specific constants.
To refine these parameters, we measure variations in power consumption at different processor utilizations and apply linear regression.

\begin{equation}
    \label{equ:freq_power}
    P_{f} = \alpha V_{f}^2 U_{f} f + \beta  V_{f}
\end{equation}

Power consumption measurements are facilitated using a USB power meter~\cite{UM25C}. 
This USB power meter quantifies the power consumption of the entire platform, denoted as $P$, with all unused processors deactivated to exclude their impact on the total power consumption. To ascertain the influence of non-processor components on the total power, all processors are turned off, and this power measurement is labeled as $P_{non}$. The power consumption of the specific processor under test is then calculated by subtracting $P_{non}$ from $P$, represented as $P-P_{non}$.

For GPU utilization, data is sourced from kernel activity logs via sysfs attributes \cite{mochel2005sysfs}. Sysfs, a RAM-based filesystem, exports kernel data structures, attributes, and linkages to user space, thus facilitating the monitoring of hardware and system metrics. CPU cluster utilization is monitored using the \textit{htop} tool, which helps us calculate the average utilization of included cores, representing the cluster’s overall utilization.

After establishing the power estimation equation~\eqref{equ:freq_power}, we measure processor utilization to estimate average power consumption accurately. Since model operators typically do not fully utilize CPU or GPU resources, we directly measure the processor utilization across the entire model. We hypothesize that this measurement effectively represents the utilization of individual operators. Furthermore, by analyzing the processors' spare time indicated by mapping strategies and observing processor utilization during the execution of assigned operators, we can calculate the average processor utilization associated with specific mapping strategies.

\subsection{Strategy Selection}
\label{sec:strategy_select}
Following the estimation, we offer users a range of Pareto-optimal configuration options tailored to their preferences for throughput and latency trade-offs. Energy efficiency consideration of generated mapping strategies is also included as a selection metric. 

\subsubsection{Pareto Optimal Analysis}
\begin{figure}[tb]
    \raggedleft
    \begin{tikzpicture}
    \centering
    \pgfplotsset{every tick label/.append style={font=\tiny}}
    \pgfplotstableread[col sep=space,header=true]{
        x y label
        0.5798 1.73 {\em para-only}
    }\paraalldat
    \pgfplotstableread[col sep=space,header=true]{
        x y label
        1.171 2.51 {\em pipe-only}
    }\pipealldat
    \pgfplotstableread[col sep=space,header=true]{
        x y label
        0.815066 2.434 {\em hybrid-T}
    }\parapipethrdat
    \pgfplotstableread[col sep=space,header=true]{
        x y label
        0.617437 2.023 {\em hybrid-L}
    }\parapipelatdat
    \begin{groupplot}[
        group style={
        group name=my plots,
        group size=1 by 1,
        xlabels at=edge bottom,
        xticklabels at=edge bottom,
        vertical sep=0pt,
        horizontal sep=0pt,},
        x tick label style={font=\tiny,rotate=0,text height=0.1pt,text width=1.5cm,align=center,inner sep=3pt,outer sep=2pt},
        y tick label style={font=\tiny,rotate=0,text height=0.1pt,inner sep=0pt,outer sep=1.5pt},
        xlabel style={font=\scriptsize, align=center, text height=0pt,inner sep=2pt,outer sep=2pt},
        ylabel style={font=\scriptsize, align=center, text height=0pt,inner sep=0pt,outer sep=2pt},]

        \nextgroupplot[width=1\columnwidth, height=3.8cm,
                        legend entries={All Options,Pareto-optimal Options,Balanced Options},
                        xlabel={Latency (s)},
                        ylabel={Throughput (fps)},
                        ymin=1.3,
                        ymax=3,
                        xmin=0.55,
                        xmax=1.5,
                        xmode=log,
                        ymode=log,
                        xtick={0.6,0.7,0.8,0.9,1,1.1,1.2,1.3,1.4,1.5,1.6},
                        xticklabels={0.6,0.7,0.8,0.9,1,1.1,1.2,1.3,1.4,1.5,1.6},
                        ytick={1.5,2,2.5},
                        yticklabels={1.5,2,2.5},
                        xmajorgrids=true,
                        ymajorgrids=true,
                        grid style=densely dashed,
                        legend cell align={left},
                        legend style={at={(0.96,1.25)},draw=none,font=\scriptsize, fill opacity=0.85, text opacity = 1, legend columns=3,cells={align=center}, column sep=0.05cm},anchor=north,]

        \addlegendimage{only marks, mark size=2pt, mark=pentagon*, timedOutPointsColor,mark options={line width=0.1pt, draw=deeperGreen,solid},}
        \addlegendimage{only marks, mark size=3pt, mark=otimes, darkgoldenrod,mark options={line width=0.5pt,solid},}
        \addlegendimage{only marks, mark size=3pt, mark=otimes, deeperBlue,mark options={line width=0.5pt,solid},}

        \addplot+[scatter, only marks, mark size=2.5pt, mark=otimes, mark options={line width=0.5pt,solid},
          scatter/classes={
                           No={mark size=0pt},
                           Yes={darkgoldenrod}, 
                           Trade={deeperBlue}}, 
          scatter src=explicit symbolic
          ] 
          table[y index=0,x index=1, meta index=2]{data/parapipe_resnetv2.dat};

        \addplot+[scatter, only marks, mark size=1.4pt, mark=pentagon*,mark options=solid,mark options={line width=0.1pt, draw=deeperGreen}, 
          scatter/classes={No={fill=timedOutPointsColor},
                           Yes={fill=timedOutPointsColor},
                           Trade={fill=timedOutPointsColor}},
          scatter src=explicit symbolic
          ] 
          table[y index=0,x index=1, meta index=2]{data/parapipe_resnetv2.dat};
        
        \addplot[darkgoldenrod,mark=otimes,mark options={fill=darkgoldenrod,line width=0.5pt,solid},nodes near coords,only marks,
            point meta=explicit symbolic,mark size=2.5pt,
            visualization depends on={value \thisrow{label} \as \Label},
            every node near coord/.append style={anchor=west,font=\scriptsize\bf,color=darkgoldenrod,xshift=-1em,yshift=-0.7em}]
            table[meta=label] \paraalldat;    
        \addplot[deeperBlue,mark=pentagon*,mark options={line width=0.1pt, draw=deeperGreen,fill=timedOutPointsColor},only marks,
            point meta=explicit symbolic,mark size=1.4pt]
            table[meta=label] \paraalldat;
   
        \addplot[darkgoldenrod,mark=otimes,mark options={fill=darkgoldenrod,line width=0.5pt,solid},nodes near coords,only marks,
               point meta=explicit symbolic,mark size=2.5pt,
               visualization depends on={value \thisrow{label} \as \Label},
               every node near coord/.append style={anchor=north,font=\scriptsize\bf,color=darkgoldenrod,yshift=0em,xshift=-0.4em}]
                 table[meta=label] \pipealldat;
        \addplot[deeperBlue,mark=pentagon*,mark options={line width=0.1pt, draw=deeperGreen,fill=timedOutPointsColor},only marks,
            point meta=explicit symbolic,mark size=1.4pt]
            table[meta=label] \pipealldat;

        \addplot[darkgoldenrod,mark=otimes,mark options={fill=darkgoldenrod,line width=0.5pt,solid},nodes near coords,only marks,
               point meta=explicit symbolic,mark size=2.5pt,
               visualization depends on={value \thisrow{label} \as \Label},
               every node near coord/.append style={anchor=south,font=\scriptsize\bf,color=darkgoldenrod,yshift=0em,xshift=0em}]
                 table[meta=label] \parapipethrdat;
        \addplot[deeperBlue,mark=pentagon*,mark options={line width=0.1pt, draw=deeperGreen,fill=timedOutPointsColor},only marks,
            point meta=explicit symbolic,mark size=1.4pt]
            table[meta=label] \parapipethrdat;

        \addplot[darkgoldenrod,mark=otimes,mark options={fill=darkgoldenrod,line width=0.5pt,solid},nodes near coords,only marks,
               point meta=explicit symbolic,mark size=2.5pt,
               visualization depends on={value \thisrow{label} \as \Label},
               every node near coord/.append style={anchor=north,font=\scriptsize\bf,color=darkgoldenrod,yshift=0.1em,xshift=0.5em}]
                 table[meta=label] \parapipelatdat;
        \addplot[deeperBlue,mark=pentagon*,mark options={line width=0.1pt, draw=deeperGreen,fill=timedOutPointsColor},only marks,
            point meta=explicit symbolic,mark size=1.4pt]
            table[meta=label] \parapipelatdat;

    \end{groupplot}  
\end{tikzpicture}
    \caption{Pareto optimal analysis of mapping options in {\em Para-Pipe} for Inception-ResNet-v2 on Amlogic SoC: Balanced options are strategies achieving moderate throughput and latency performance.}
    \label{fig:latency-throughput-trade-off}
\end{figure}
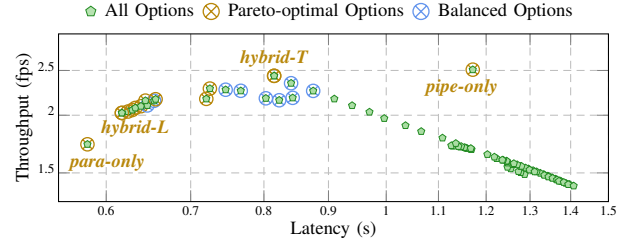

Based on estimated latency and throughput, configurations exhibiting inferior performance across both metrics are excluded, while those demonstrating Pareto optimality are retained. Fig.~\ref{fig:latency-throughput-trade-off} illustrates the performance trade-offs associated with various \textit{Para-Pipe} configurations for Inception-ResNet-v2 on the Amlogic SoC, employing coarse-grained mapping to develop parallel operator strategies. 

The pipe-only configuration achieves maximum throughput but suffers from impractical latency levels. Conversely, the para-only setup minimizes latency at the expense of significantly reduced throughput. Hybrid configurations present multiple Pareto-optimal solutions that effectively balance throughput with latency. When both metrics are accorded equal importance (Balanced Options), {\em Para-Pipe} also yields several commendable configurations for users to choose from.
This study focuses on pipe-only, para-only, and hybrid options that achieve the lowest latency and highest throughput (hybrid-L and hybrid-T) for further analysis.

\subsubsection{Estimation Accuracy}
\begin{figure}[tb]
    \raggedleft
    \pgfplotsset{every axis/.append style={
    line width=0.05pt,
    tick style={line width=0.6pt},
    mark options={line width=0.9pt, solid},
    },
    }

\begin{tikzpicture}
    \begin{groupplot}[
        group style={
        group name=my plots,
        group size=2 by 1,
        xlabels at=edge bottom,
        xticklabels at=edge bottom,
        vertical sep=0pt,
        horizontal sep=35pt,},
        xtick=data,
        x tick label style={font=\scriptsize,rotate=15,text height=0pt,text width=1.5cm,align=center,inner sep=2pt,outer sep=2pt},
        xticklabels from table={data/parapipe_energy_prediction.dat}{Model},
        xlabel style={font=\scriptsize, align=center, text height=0pt,inner sep=0pt,outer sep=0pt},
        ytick pos=left,
        y tick label style={font=\tiny,rotate=0,text height=0pt,inner sep=1.5pt,outer sep=0pt},
        ylabel style={font=\scriptsize, align=center, text height=0.1pt,inner sep=0pt,outer sep=0pt},]

        \nextgroupplot[width=0.54\columnwidth,, height=3.6cm, 
                        ylabel={Throughput Error (\%)},
                        ymin=-35,
                        ymax=35,
                        ymajorgrids=true,
                        grid style=dotted,
                        enlarge x limits={abs=0.5},
                        ytick={-30,-20,-10,0,10,20,30},
                        yticklabels={-30,-20,-10,0,10,20,30},
                        extra y ticks={0},
                        extra tick style={grid=major,major grid style={red, densely dashed}},
                        extra y tick labels={},
                        typeset ticklabels with strut,
                        legend columns=6,
                        legend style={at={(1.18,1.11,1.15)}, anchor=south, font=\scriptsize}]
    
        \addplot+[mark=square, slategray, mark size=1.2pt,]table[x expr = \coordindex,y = para-only,]{data/parapipe_throughput_prediction.dat};
        \addlegendentry{\em para-only}

        \addplot+[mark=x, darkgoldenrod, mark size=2.2pt]table[x expr = \coordindex,y = hybrid-L,]{data/parapipe_throughput_prediction.dat};
        \addlegendentry{\em hybrid-L}

        \addplot+[mark=+, deeperGreen, mark size=2.2pt]table[x expr = \coordindex,y = hybrid-T,]{data/parapipe_throughput_prediction.dat};
        \addlegendentry{\em hybrid-T}

        \addplot+[mark=diamond, deeperBlue, mark size=1.8pt]table[x expr = \coordindex,y = pipe-only,]{data/parapipe_throughput_prediction.dat};
        \addlegendentry{\em pipe-only}

        \draw (1,-25) node[align=center] {\scriptsize \bfseries \textcolor{red}{RMSPE: 15.25\%}};

        \nextgroupplot[width=0.54\columnwidth, height=3.6cm,
                        ylabel={Energy Efficiency\\Error (\%)},
                        ylabel style={font=\scriptsize, align=center, text height=0pt},
                        xlabel style={font=\scriptsize, align=center, text height=0pt},
                        ymin=-35,
                        ymax=35,
                        ymajorgrids=true,
                        grid style=dotted,
                        enlarge x limits={abs=0.5},
                        ytick={-30,-20,-10,0,10,20,30},
                        yticklabels={-30,-20,-10,0,10,20,30},
                        extra y ticks={0},
                        extra tick style={grid=major,major grid style={red, densely dashed}},
                        extra y tick labels={},
                        typeset ticklabels with strut,]
    
        \addplot+[mark=square, slategray, mark size=1.2pt,]table[x expr = \coordindex,y = para-only,]{data/parapipe_energy_prediction.dat};

        \addplot+[mark=x, darkgoldenrod, mark size=2.2pt]table[x expr = \coordindex,y = hybrid-L,]{data/parapipe_energy_prediction.dat};

        \addplot+[mark=+, deeperGreen, mark size=2.2pt]table[x expr = \coordindex,y = hybrid-T,]{data/parapipe_energy_prediction.dat};
        
        \addplot+[mark=diamond, deeperBlue, mark size=1.8pt]table[x expr = \coordindex,y = pipe-only,]{data/parapipe_energy_prediction.dat};

        \draw (1,-25) node[align=center] {\scriptsize \bfseries \textcolor{red}{RMSPE: 6.40\%}};
        
    \end{groupplot}           
\end{tikzpicture}
    \vspace*{-0.2cm}
    \caption{Prediction error of throughput (left) and energy efficiency (right) of selected \textit{Para-Pipe} mapping options on Amlogic SoC.}
    \label{fig:para-pipe-cost-prediction}
\end{figure}
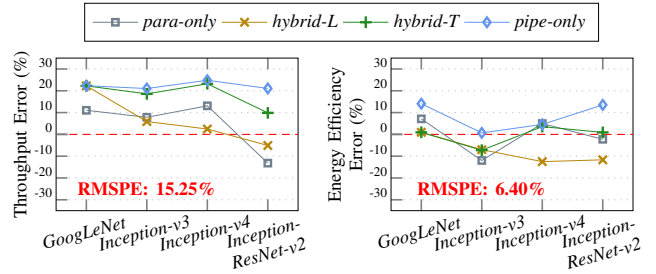

To assess the effectiveness of our cost estimator, we evaluated it across four configurations: para-only, hybrid-L, hybrid-T, and pipe-only. Our findings indicate that the root mean squared prediction error (RMSPE) for latency, throughput, and energy efficiency stands at 15.33\%, 15.25\%, and 6.40\%, respectively, across all benchmarks. Details of the throughput and energy efficiency prediction errors are illustrated in Fig.~\ref{fig:para-pipe-cost-prediction}. 
Despite an approximate 15\% prediction error in latency and throughput, the cost estimator effectively determines the relative performance rankings of mapping strategies for both metrics. These rankings have been verified on the real platform using runtime evaluations, ensuring the reliability of the cost estimator in identifying robust and practical Pareto-optimal solutions.

\section{Runtime Management on SoC}
\begin{figure}[tb]
    \centering
    \includegraphics[width=0.7\linewidth]{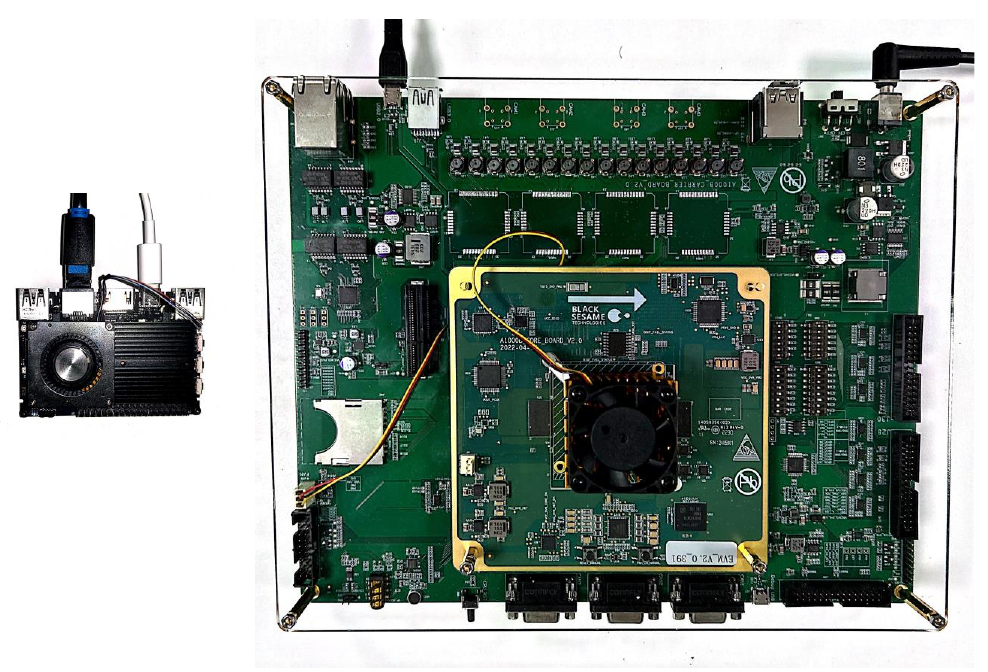}
    \caption{Picture of Khadas VIM3 Pro board (left) and BST A1000 evaluation board (right).}
    \label{fig:bst-khadas-board}
\end{figure}

\begin{figure}[t]
    \centering
    \includegraphics[width=0.7\linewidth]{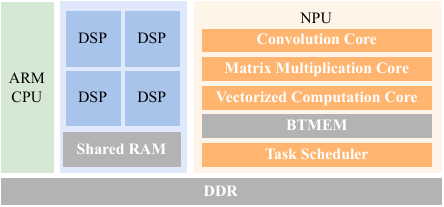}
    \caption{BST A1000 SoC hardware architectural diagram.}
    \label{fig:bst-arc}
\end{figure}

We implement the runtime management of \textit{Para-Pipe} onto \textit{Amlogic A311D SoC of Khadas Vim3 Pro Platform}~\cite{Khadas_Vim3_Pro} (Fig.~\ref{fig:bst-khadas-board}). Additionally, we conduct performance estimation using the \textit{BST A1000 evaluation board} (Fig.~\ref{fig:bst-arc}) provided by~\cite{BST_Company}, which is explained in detail in Section~\ref{subsec:bst}.

\subsection{Implementation on Amlogic SoC}
\label{sec:amlogic_runtime}
The Amlogic SoC incorporates an \textit{ARM big.LITTLE} CPU architecture, featuring a high-performance quad-core Cortex-A73 cluster and a power-efficient dual-core Cortex-A53 cluster, along with an \textit{ARM} G52 MP4 GPU, as depicted in Fig.~\ref{fig:Para-Pipe}.
Since most ML libraries optimize at the operator level for sequential neural networks, {\em Para-Pipe} can be adapted to any of these libraries to achieve specific latency-throughput trade-offs and improved energy efficiency without compromising operator-level optimization. In this paper, we build the {\em Para-Pipe} runtime for Amlogic SoC on top of \textit{ARM Compute Library (ARM-CL)}, which provides a collection of state-of-the-art low-level ML functions optimized for \textit{ARM} Cortex-A CPU and Mali GPU architectures. \textit{ARM-CL} features a \textit{Graph} API for constructing a model graph representation employing graph-level optimizations to enhance efficiency. This library employs \textit{Neon} (or \textit{SVE}) for acceleration on \textit{ARM} CPUs and \textit{OpenCL} for GPU.

\begin{figure}[tb]
    \centering
    \includegraphics[width=0.7\linewidth]{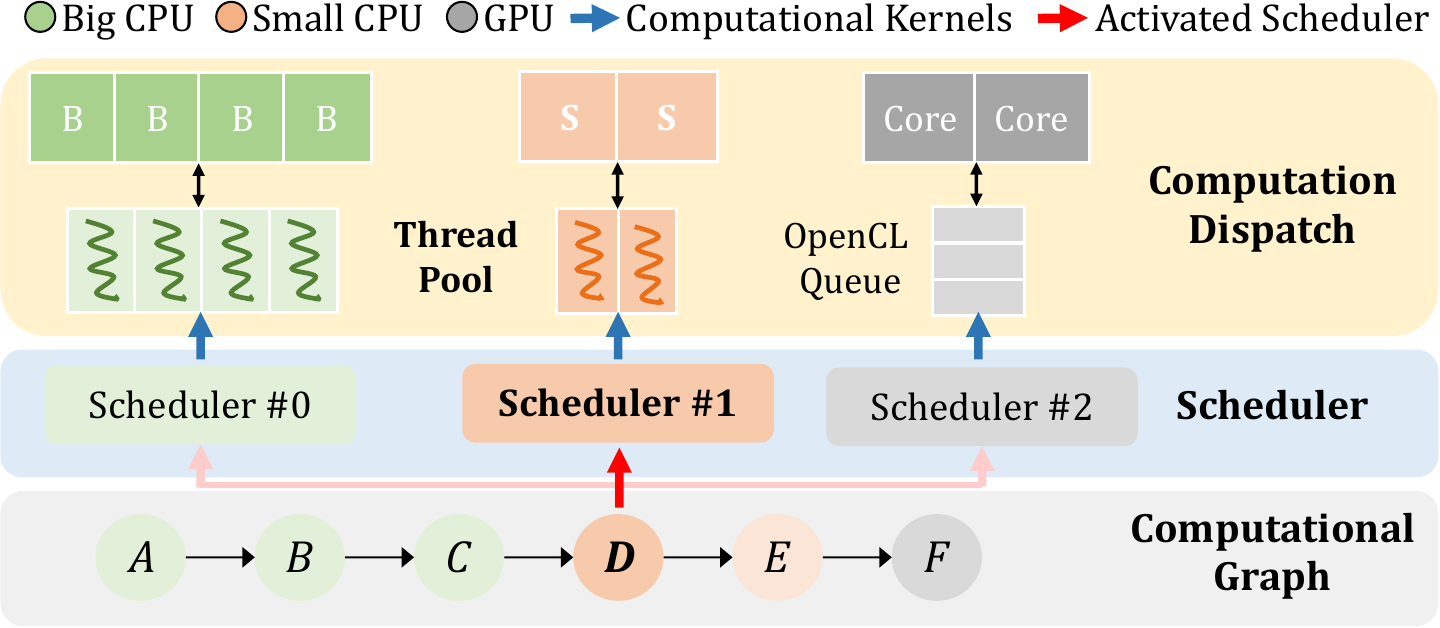}
    \caption{\textit{ARM-CL} runtime illustration: sequential execution of computational graph across big CPU cluster, small CPU cluster, and GPU on Amlogic SoC.}
    \label{fig:amlogic_runtime}
\end{figure}

Since the default \textit{ARM-CL} only supports sequential execution onto CPUs or GPU, we configure unique target backends for CPUs and GPU to distribute graph nodes across processors. Nodes are assigned to specific backends based on our mapping strategy. We employ separate schedulers for CPU clusters and GPU to support the concurrent execution of operators; a simplified illustration of the sequential execution of operators across processors is depicted in Fig.~\ref{fig:amlogic_runtime}. The schedulers for CPUs are associated with a thread pool consisting of worker threads bound to the corresponding CPU cluster cores. During node execution, the scheduler pertinent to the target backend is invoked. The CPU scheduler manages the execution by dispatching computation kernels to worker threads, while the GPU scheduler enqueues computational kernels for asynchronous execution in the \textit{OpenCL} queue.

\begin{figure}[tb]
    \centering
    \includegraphics[width=0.99\linewidth]{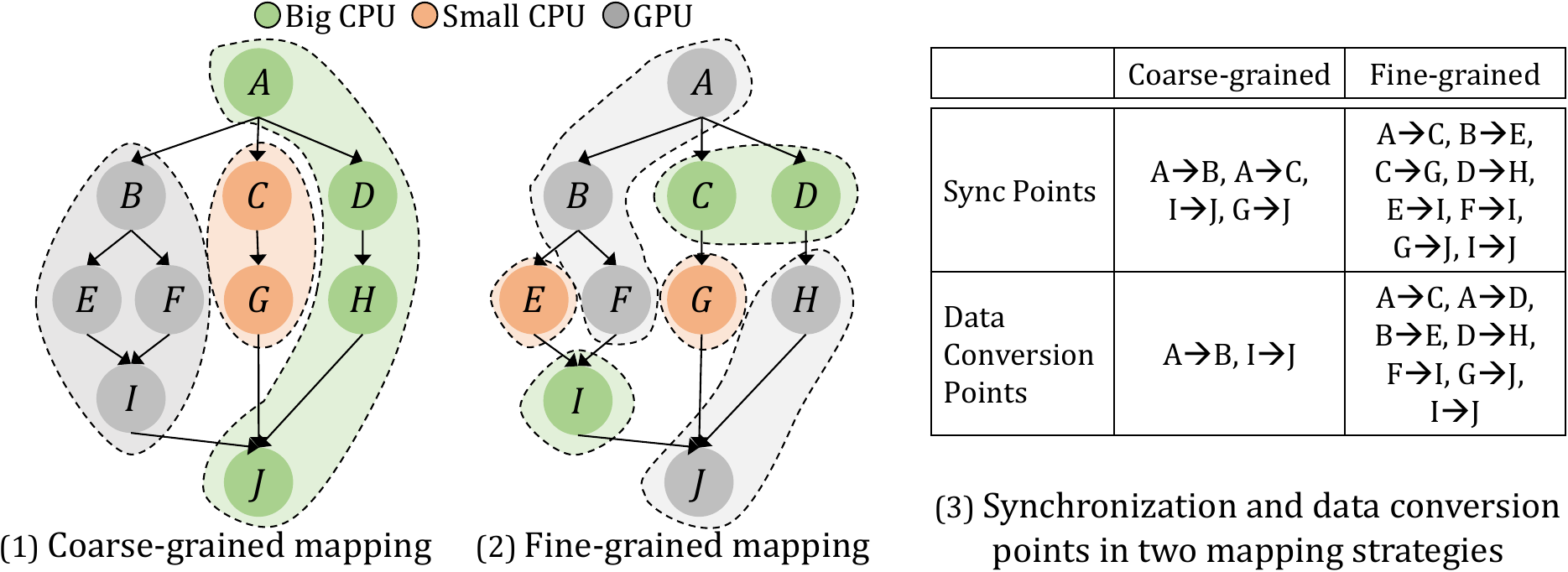}
    \caption{Runtime comparison of coarse- vs. fine-grained mapping strategies: Analyzing synchronization and inter-processor communication overhead.}
    \label{fig:sync_impl}  
\end{figure}

\subsubsection{Pipeline Deployment} We generate an \textit{ARM-CL} graph for each pipeline stage and execute it on a single thread. In cases where multiple engines are utilized for computation within a stage, the same number of threads are initiated for independent execution. 
Meanwhile, a data buffer stores transferred data between consecutive stages. Whenever a stage produces output data, its successor stage is notified. 

\subsubsection{Co-execution of CPUs and GPU}
\label{subsubsec:co-exec}
Separate threads are spawned to concurrently execute nodes on both CPU clusters and GPUs, with semaphores employed to manage data dependencies. Regarding inter-node data transfer between CPU clusters, no data conversion is required. However, data conversion is imperative for the intricate communication between CPU clusters and GPU.
GPU computation exclusively operates on data formatted in the \textit{OpenCL} tensor format. When GPU computation requires CPU data, a specialized \textit{OpenCL} tensor is created to accommodate the transferred data. Conversely, an additional address mapping is required to facilitate CPU access to GPU output within the device address space.
Fig.~\ref{fig:sync_impl} compares the runtime overhead of coarse- and fine-grained mapping strategies.  The comparison indicates that a coarse-grained approach exhibits reduced susceptibility to synchronization overhead and frequent communication demands.

\subsection{Implementation on BST SoC}
\label{subsec:bst}
The BST A1000 SoC offers high-performance, low-power neural network inference. It incorporates an ASIC accelerator NPU and four Cadence DSP cores. It is designed for a wide range of embedded applications, with a central focus on camera-based sensing and AI computing for Level-3 autonomous driving systems. 
Fig.~\ref{fig:bst-arc} provides a simplified view of the BST A1000's hardware architecture. 
In real-world use cases, two DSPs are reserved for other uses in autonomous driving. Therefore, we use a 1 NPU + 2 DSP combination to simulate a realistic scenario. 

We utilize an operator simulator provided by~\cite{BST_Company} to estimate runtimes on both processing engines and the communication times between them. The runtime and communication cost estimation for operators utilizes historical data from previously tested configurations as a benchmark. For operators without prior testing under diverse parameter settings, we apply a k-means clustering algorithm to predict runtime and communication cost based on operator similarities. These estimations serve as the foundation for applying our {\em Para-Pipe} framework, providing data for the ILP formulation in the Parallel Operator Mapper and enabling the estimation of latency and throughput for the generated mapping strategies.
\section{Experimental Evaluation}
This section evaluates the \textit{Para-Pipe} framework on Amlogic and BST SoCs, focusing on selected strategies, including pipe-only, para-only, and hybrid options targeting the lowest latency (hybrid-L) and the highest throughput (hybrid-T).
All experimental data for the Amlogic SoC, encompassing latency, throughput, and energy efficiency, were obtained from the real Khadas Vim3 Pro platform. Conversely, the data for the BST SoC are derived from simulations.

\subsection{Experimental Setup}

\setlength\tabcolsep{2pt}
\begin{table}[t]
    \centering
    \caption{Structural details of benchmark models}
    \label{tab:used_benchmarks}
    \resizebox{\linewidth}{!}{
        \begin{tabular}{|c|c|c|c|}
        \hline
        Network&\makecell{\#Subgraphs}&\makecell{Maximum\\Parallelism}&Major Operators / Modules\\
            \hline
            \hline
            GoogLeNet&11&4&\makecell[c]{3 Conv + 9 Inception Modules + 1 FC}\\
            \hline
            Inception-v3&13&6&\makecell[c]{5 Conv + 11 Inception Modules + 1 FC}\\
            \hline
            Inception-v4&21&6&\makecell[c]{11 Conv + 14 Inception Modules\\+ 2 Reduction Modules + 1 FC}\\
            \hline
            \makecell{Inception-\\ResNet-v2}&45&4&\makecell[c]{13 Conv + 40 Residual Inception\\Modules + 2 Reduction Modules + 1 FC}\\
            \hline
            \makecell{PETR-based}& 24 & 16 &\makecell[c]{ResNet-50 + 3D Position Encoder +\\Transformer Decoder + Query Generator}\\
            \hline
            \makecell{BEVFormer-\\based}& 25 & 4 &\makecell[c]{6 Spatial Cross-attention + 6\\Transformer + 6 Temporal Self-attention}\\
        \hline
        \end{tabular}
    }
\end{table}
\subsubsection{Benchmarks}
The pipelined execution of linear models was comprehensively evaluated in our prior work~\cite{aghapour2023pipelined} on the Amlogic SoC. This study focuses on six modern models with dense operator parallelism to evaluate the performance of {\em Para-Pipe}.
The chosen models, GoogLeNet, Inception-v3, Inception-v4~\cite{szegedy2017inception}, and Inception-ResNet-v2~\cite{szegedy2017inception} exemplify conventional regular DNNs with distinct independent branch structures. 
Two networks featuring complex and irregular connections, i.e., PETR-based network~\cite{liu2022petr} and BEVFormer-based~\cite{li2022bevformer}, are the state-of-the-art in autonomous driving. Some operators included are not supported by \textit{ARM-CL} on Amlogic SoC; so we evaluate them on BST SoC to demonstrate the platform portability of our framework. Table~\ref{tab:used_benchmarks} outlines model structural details, including the number of subgraphs generated by the Graph Partitioner, serving as partition points in \textit{Para-Pipe} Pipeline Configuration Generator.

\subsubsection{Platform Setup}
We assess the performance and efficiency of the Amlogic SoC using a real Khadas Vim3 Pro platform. In our experiments, the platform's big and small CPU clusters are set to operate at their maximum voltage and frequency levels—1.04 V/2.2 GHz and 0.83 V/1.8 GHz, respectively. Additionally, the \textit{ARM} GPU on the platform is configured to run at its peak performance level of 1.15 V/0.8 GHz. We connect the board to a host machine via an Ethernet cable using Secure Shell (SSH) for reliable and secure data transfer. A Khadas-provided DC 5 V cooling fan is utilized throughout testing to mitigate potential thermal instabilities.

We input a continuous stream of 50 frames for each mapping strategy and measure the average throughput, quantified as frames processed per second and the average latency per frame. A USB power meter~\cite{UM25C} is employed to monitor the total power consumption of the board, from which we calculate the average active power and energy efficiency. 
Following each run, the board is left idle to cool down adequately before proceeding with the next test.

\subsubsection{Baseline}
In this study, we establish two baseline configurations. Firstly, we employ the Layer-switched algorithm~\cite{aghapour2022cpu} to illustrate the performance capabilities of traditional sequential execution of models utilizing all on-chip processor resources. The Layer-switched algorithm supports layer-wise transitions among processors and follows a sequential execution strategy to minimize latency by effectively managing the switching overhead between processors. 
Notably, the algorithm achieves superior latency and throughput performance without leveraging operator or pipeline parallelism, thus serving as an upper bound for sequential execution performance when utilizing any single or multiple hardware platform components.
This evaluation on Amlogic SoC utilizes the tool reported in~\cite{aghapour2024arm}. 

Secondly, we implement and compare two classic DAG mapping algorithms, HEFT and CPOP, as detailed in~\cite{topcuoglu2002performance}. For each tested model, we select the algorithm that demonstrates optimal latency to serve as a comparative baseline in our experiments, and we refer to this algorithm as HEFT \& CPOP throughout this paper.
HEFT algorithm prioritizes tasks based on the highest upward rank value at each decision point, assigning tasks to the processor that enables the earliest completion. In contrast, CPOP integrates both upward and downward ranks to prioritize tasks, specifically scheduling those on the critical path to the processor that minimizes total execution time. We use these algorithms to evaluate our parallel operator mapping algorithms rigorously.

\subsection{Resultant Configurations}
This section briefly analyzes the pipelined configurations of the hybrid-L and hybrid-T options in {\em Para-Pipe}. On Amlogic SoC, both options utilize a dual-stage configuration employing CPU clusters and the GPU across all supported models. The configuration utilizes a coarse-grained algorithm for parallel operator mapping within the stages. 
This is because parallel processing within a single stage prefers processors with similar architectures and data formats to reduce unnecessary data conversion and communication overhead. The details are further explained in Section~\ref{subsec:co-diverse-units}. 
Conversely, the partial operator support of BST SoC's NPU, necessitating the offloading of certain operators to DSP, limits the feasibility of \textit{Para-Pipe}'s ideal operator or stage arrangement. Therefore, for most benchmark models, the hybrid options mostly combine NPU and DSP in one stage while solely using DSP in another. The DSP is leveraged not only as a fallback for operators unsupported by the NPU but also to exploit operator parallelism within subgraphs when feasible. The irregular models evaluated on BST SoC necessitate a fine-grained mapping approach.

\setlength\tabcolsep{3pt}
\begin{table}[tb]
    \centering
    \caption{Subgraph allocations for hybrid-L and hybrid-T strategies across four representative models in {\em Para-Pipe} on BST SoC}
    \label{tab:para-pipe_subgraph_alloc_config}
        \begin{tabular}{|l|l|l|l|}
        \hline
            \multicolumn{1}{|c|}{Network}&Option&Pipeline Config.&Subgraph Allocation\\
            \hline
            \hline
            \multirow{2}{*}{GoogLeNet}& hybrid-L & NPU - 2DSPs & [1,10] - [11] \\
                \cline{2-4}  
                & hybrid-T & NPU - 2DSPs & [1,9] - [10,11] \\
            \hline
            \multirow{2}{*}{Inception-v4}&hybrid-L&NPU+DSP - DSP&[1,20] - [21]\\
                \cline{2-4} 
                &hybrid-T&DSP - NPU+DSP&[1,3] - [4,21]\\
            \hline
            \multirow{2}{*}{PETR-based}&hybrid-L&NPU+DSP - DSP&[1,17] - [18,24]\\
                \cline{2-4}
                &hybrid-T&NPU+DSP - DSP&[1,15] - [16,24]\\
            \hline
            \multirow{2}{*}{BEVFormer-based}&hybrid-L&NPU+DSP - DSP&[1,23] - [24,25]\\
                \cline{2-4}
                &hybrid-T&DSP - NPU+DSP&[1,2] - [3,25]\\
        \hline
        \end{tabular}
\end{table}

To understand how {\em Para-Pipe} selects pipeline structures with preferences for latency and throughput, we focus on the BST SoC and select four representative benchmarks for enhanced visualization. The allocations for the hybrid-L and hybrid-T strategies are detailed in Table~\ref{tab:para-pipe_subgraph_alloc_config}. The hybrid-T option typically maintains a balanced pipeline configuration, as it enhances throughput by decreasing delays in the longest stage while concurrently distributing a significant workload to other stages.
Based on the data in Table~\ref{tab:para-pipe_subgraph_alloc_config}, we identify two methods by which the hybrid-L option accelerates execution per frame relative to the hybrid-T option. The first method maps more subgraphs onto the high-performance NPU, decreasing the overall execution time for a single frame, as demonstrated in the GoogLeNet configuration. The second method boosts intra-stage operator parallelism by allocating additional subgraphs for parallel processing, as seen in the configurations for remaining networks. {\em Para-Pipe} can adaptively adjust workloads within and across stages to optimize latency and throughput trade-offs through hierarchical mapping passes.

\begin{figure*}[!t]
    \raggedleft
        \input{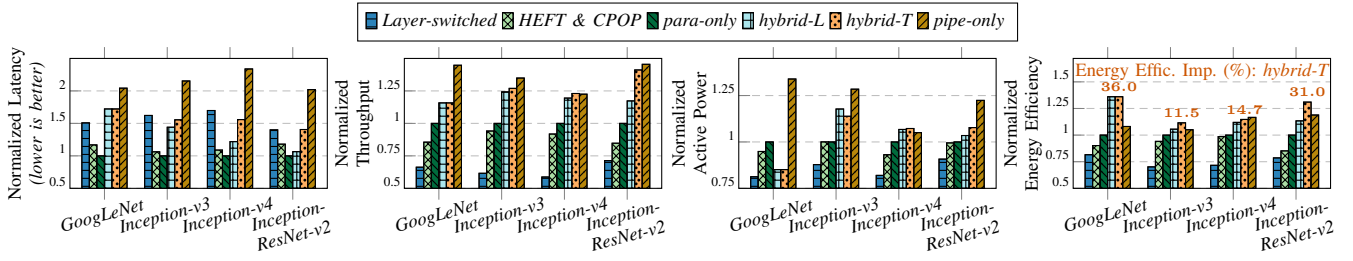}
        \vspace*{-0.1cm}
        \caption{Normalized latency, throughput, active platform power, and energy efficiency for Layer-switched method, HEFT \& CPOP algorithm and four representative {\em Para-Pipe} mapping strategies on Amlogic SoC relative to the para-only option.}
        \label{fig:para-pipe-throughput-latency-energy-normalized}
\end{figure*}

\subsection{Latency-Throughput Trade-off}
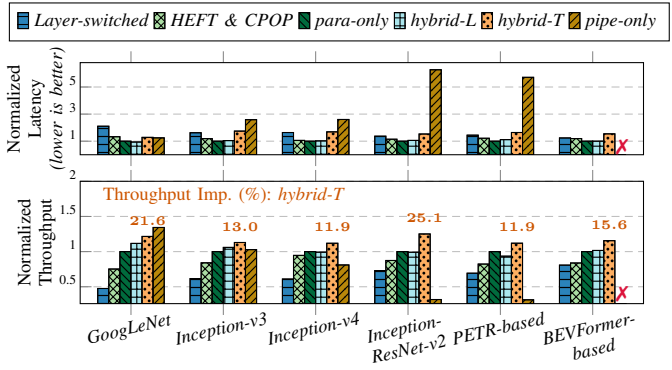
\begin{figure}[tb]
    \begin{tikzpicture}
    \centering
    \begin{groupplot}[
        group style={
            group name=my plots,
            group size=1 by 2,
            xlabels at=edge bottom,
            xticklabels at=edge bottom,
            vertical sep=10pt,
            horizontal sep=0pt,},
        xtick=data,
        x tick label style={font=\scriptsize,rotate=13,text height=0pt,text width=1.5cm,align=center,inner sep=1pt,outer sep=1pt},
        xticklabels from table={data/latency_para_pipe_bst_normalized.dat}{Model},
        xlabel style={font=\scriptsize, align=center, text height=0pt,inner sep=0pt,outer sep=0pt},
        ytick pos=left,
        y tick label style={font=\tiny,rotate=0,text height=0pt,inner sep=1.5pt,outer sep=0pt},
        ylabel style={font=\scriptsize, align=center, text height=0.1pt,inner sep=0pt,outer sep=0pt},]

        \nextgroupplot[width=1.02\columnwidth, height=2.8cm,
                        ylabel={Normalized\\Latency\\\textit{(lower is better)}},
                        ylabel style={font=\scriptsize, align=center, text height=0pt},
                        xlabel style={font=\scriptsize, align=center, text height=0pt},
                        bar width=0.12,
                        ybar=0pt,    
                        ymin=0,
                        ymax=6.8,
                        ymajorgrids=true,
                        enlarge x limits={abs=0.55},
                        ytick={1,3,5,7},
                        yticklabels={1,3,5,7},
                        grid style=densely dashed,
                        typeset ticklabels with strut,
                        legend columns=6,
                        legend style={at={(0.46,1.23,1.25)}, anchor=south, font=\scriptsize}]

        \addplot[fill=rendering,postaction={pattern=horizontal lines}]
        table[x expr = \coordindex, y = layer-switched,]{data/latency_para_pipe_bst_normalized.dat};
        \addlegendentry{\em Layer-switched} 

        \addplot[fill=timedOutPointsColor,postaction={pattern=crosshatch}]
        table[x expr = \coordindex, y = heft,]{data/latency_para_pipe_bst_normalized.dat};
        \addlegendentry{\em HEFT \& CPOP} 
        
        \addplot[fill=cadmiumgreen,postaction={pattern=north west lines}]
        table[x expr = \coordindex, y = para-only,]{data/latency_para_pipe_bst_normalized.dat};
        \addlegendentry{\em para-only}

        \addplot[fill=blizzardblue,postaction={pattern=grid}]
        table[x expr = \coordindex, y = hybrid-L,]{data/latency_para_pipe_bst_normalized.dat};
        \addlegendentry{\em hybrid-L}

        \addplot[fill=unsatPointsColor,postaction={pattern=crosshatch dots},
        ]
        table[x expr = \coordindex, y = hybrid-T,]{data/latency_para_pipe_bst_normalized.dat};
        \addlegendentry{\em hybrid-T}

        \addplot[fill=darkgoldenrod,postaction={pattern=north east lines}]
        table[x expr = \coordindex, y = pipe-only,]{data/latency_para_pipe_bst_normalized.dat};
        \addlegendentry{\em pipe-only}

        \node at (5.32,0.6) {\footnotesize\xmark};

        \nextgroupplot[width=1.02\columnwidth, height=3.2cm,
                        ylabel={Normalized\\Throughput},
                        ylabel style={font=\scriptsize, align=center, text height=0pt},
                        xlabel style={font=\scriptsize, align=center, text height=0pt},
                        bar width=0.12,
                        ybar=0pt,    
                        ymin=0.26,
                        ymax=2,
                        ymajorgrids=true,
                        enlarge x limits={abs=0.55},
                        ytick={0,0.5,1,1.5,2},
                        yticklabels={0,0.5,1,1.5,2},
                        grid style=densely dashed,
                        typeset ticklabels with strut,]

        \addplot[fill=rendering,postaction={pattern=horizontal lines}]
        table[x expr = \coordindex, y = layer-switched,]{data/throughput_para_pipe_bst_normalized.dat};

        \addplot[fill=timedOutPointsColor,postaction={pattern=crosshatch}]
        table[x expr = \coordindex, y = heft,]{data/throughput_para_pipe_bst_normalized.dat};

        \addplot[fill=cadmiumgreen,postaction={pattern=north west lines}]
        table[x expr = \coordindex, y = para-only,]{data/throughput_para_pipe_bst_normalized.dat};

        \addplot[fill=blizzardblue,postaction={pattern=grid}]
        table[x expr = \coordindex, y = hybrid-L,]{data/throughput_para_pipe_bst_normalized.dat};

        \addplot[fill=unsatPointsColor,postaction={pattern=crosshatch dots}
        ,nodes near coords={\pgfkeys{/pgf/fpu}\pgfmathparse{\pgfplotspointmeta*100-100}\pgfmathprintnumber{\pgfmathresult}},nodes near coords style = { /pgf/number format/.cd, fixed, fixed zerofill, precision=1, /tikz/.cd, font=\tiny\boldmath, color=deeperorange, yshift=0.2ex, xshift=0ex},
        ]
        table[x expr = \coordindex, y = hybrid-T,]{data/throughput_para_pipe_bst_normalized.dat};

        \addplot[fill=darkgoldenrod,postaction={pattern=north east lines}]
        table[x expr = \coordindex, y = pipe-only,]{data/throughput_para_pipe_bst_normalized.dat};

        \node at (5.32,0.4) {\footnotesize\xmark};
        \draw (1,1.8) node[align=center] {\scriptsize \textcolor{deeperorange}{Throughput Imp. (\%): {\em hybrid-T}}};
        
    \end{groupplot}           
\end{tikzpicture}
    \centering
    \vspace*{-5mm}
    \caption{Normalized latency and throughput for Layer-switched method, HEFT \& CPOP algorithm and four representative {\em Para-Pipe} mapping strategies on BST SoC relative to the para-only option.}
    \label{fig:bst_lat_thr_comp}
\end{figure}

We normalize these metrics relative to the para-only method to explore the balance between latency and throughput. The comparative results for Amlogic SoC are depicted in Fig.~\ref{fig:para-pipe-throughput-latency-energy-normalized} and for BST SoC in Fig.~\ref{fig:bst_lat_thr_comp}.

Our findings indicate that traditional sequential execution modes fail to fully utilize on-chip computing resources even with layer-wise transitions among processors when models exhibit intensive operator parallelism. The Layer-switched algorithm, although effective, cannot surpass the performance of purely parallel operator execution.

Our parallel operator mapping algorithm (para-only) significantly improves over the better-performing baseline algorithm, either HEFT or CPOP. Specifically, it achieves an average latency improvement of 10.9\% on Amlogic SoC and 15.5\% on BST SoC, and throughput improvements of 12.5\% and 18.8\%, respectively. Accurate profiling of operator runtime statistics, combined with ILP-based techniques for solving parallel mapping problems, empowers our algorithm to devise strategies that optimize for minimal execution time.

However, purely parallel operator execution does not necessarily yield higher throughput than pipelined execution, while pipelined execution inevitably causes a higher overhead in latency. 
For example, on Amlogic SoC, the pipe-only configuration approaches the highest throughput but results in an average of 113.8\% increase in latency compared to the para-only setup. Conversely, the para-only setup achieves the shortest execution times, albeit with an average of 26.6\% reduction in throughput. Hybrid strategies that prioritize either throughput or latency offer a viable compromise. The hybrid-L configuration, for instance, decreases throughput by 12.4\% but significantly reduces latency by 36.0\% compared to the pipe-only approach. Similarly, the hybrid-T strategy decreases throughput by 7.3\% but improves latency by 26.8\%. In the case of Inception-v4, the hybrid-T strategy excels, surpassing the pipe-only configuration in throughput while halving the latency.

\subsection{General Applicability}
Graph-aware modulation of intra- and inter-stage operator parallelism enables {\em Para-Pipe} to effectively navigate platform constraints, such as limited processor support for certain operators, by offering various mapping strategies. 
Here, we utilize BST SoC to illustrate this general applicability of our framework {\em Para-Pipe}. 
As shown in Fig.~\ref{fig:bst_lat_thr_comp}, limited processor support significantly undermines the efficacy of pipe-only optimizations in throughput. Specifically, unsupported operators may be situated at crucial partition points, rendering an ideal pipe-only configuration infeasible.
Compared to the para-only strategy, the pipe-only option results in substantial latency overheads without providing throughput benefits for most models. For instance, in the PETR-based model, the pipe-only configuration incurs a latency that is 5.7 times greater and a throughput that is only 31.4\% of that achieved with the para-only method. Moreover, generating an efficient pipe-only configuration for a BEVFormer-based network is infeasible.
Conversely, the hybrid-T option, which utilizes one DSP to support operators that the NPU alone cannot, during the parallel processing stage, achieves an average throughput improvement of 16.2\% across all tested models relative to the para-only strategy. This represents the highest throughput enhancement across the majority of tested models. 
Regarding latency, the hybrid-L and para-only options perform comparably. These results verify that {\em Para-Pipe} can successfully recommend adaptive mapping strategies, considering platform compatibility, to align with users’ preferences.


\subsection{Energy Efficiency}
This section uses Amlogic SoC to evaluate the energy efficiency of {\em Para-Pipe}. 
Fig.~\ref{fig:para-pipe-throughput-latency-energy-normalized} showcases the normalized average active platform power and energy efficiency relative to the para-only strategy. 
Our analysis reveals that the Layer-switched and HEFT \& CPOP algorithms exhibit average degradations of 24.5\% and 8.0\% in energy efficiency, respectively, compared to para-only. 
These declines are primarily attributed to frequent processor switching, a scheduling bias toward high-performance but less energy-efficient components, and additional communication overhead due to suboptimal parallel operator mapping. 

Compared to HEFT \& CPOP, para-only employs optimized workload mapping strategies, which reduce synchronization contention and processor idle time. Consequently, its additional area overhead primarily results from increased runtime processor utilization rather than inefficient hardware resource consumption. Runtime measurements, averaged across tested models, indicate a modest 3.3\% increase in platform active power, improved memory access patterns, and an 11.7\% reduction in mutex contention.
These trade-offs are well-justified by the substantial performance and energy efficiency benefits, making para-only highly suitable for workloads demanding low latency and high energy efficiency.

Conversely, the pipe-only configuration consistently outperforms the para-only setup across all tested benchmarks, achieving an average improvement of 12.2\%. Additionally, the hybrid-L and hybrid-T configurations demonstrate average increases in energy efficiency of 16.7\% and 23.3\%, respectively, relative to the para-only approach. When compared to the pipe-only configuration, these gains are 8.7\% for hybrid-L and 11.0\% for hybrid-T, underscoring their enhanced performance.

An exception is observed with Inception-v4, where the hybrid-T option reaches 0.53 fps/J, slightly underperforming against the pipe-only’s 0.54 fps/J, translating to a minor setback of 1.7\%. This slight decrease in energy efficiency with Inception-v4 can be ascribed to a marginal 0.4\% increase in throughput and a 33.3\% improvement in latency offered by the hybrid-T option. To encapsulate, the hybrid-T option enhances energy efficiency and presents a nuanced trade-off in performance, particularly noticeable in models with extensive operator parallelism.

\begin{figure*}[t]
    \begin{minipage}{\linewidth}
        \raggedleft
        \begin{tikzpicture}
    \centering
    \begin{groupplot}[
        group style={
        group name=my plots,
        group size=3 by 1,
        xlabels at=edge bottom,
        xticklabels at=edge bottom,
        vertical sep=0pt,
        horizontal sep=32pt,},
        xtick=data,
        x tick label style={font=\scriptsize,rotate=13,text height=0pt,text width=1.5cm,align=center,inner sep=0pt,outer sep=0.5pt},
        xticklabels from table={data/latency_normalized.dat}{Model},
        xlabel style={font=\scriptsize, align=center, text height=0pt,inner sep=0pt,outer sep=0pt},
        ytick pos=left,
        y tick label style={font=\tiny,rotate=0,text height=0pt,inner sep=1.5pt,outer sep=0pt},
        ylabel style={font=\scriptsize, align=center, text height=0.1pt,inner sep=0pt,outer sep=0pt},]

        \nextgroupplot[width=0.355\columnwidth,, height=3.3cm,
                        ylabel={Normalized Latency\\\textit{(lower is better)}},
                        bar width=0.12,
                        ybar=0pt,    
                        ymin=0.25,
                        ymax=3.08,
                        ymajorgrids=true,
                        grid style=densely dashed,
                        enlarge x limits={abs=0.55},
                        ytick={0,0.5,1,1.5,2,2.5},
                        yticklabels={0,0.5,1,1.5,2,2.5},
                        typeset ticklabels with strut,
                        legend columns=6,
                        legend style={at={(1.76,1.13,1.25)}, anchor=south, font=\scriptsize}]
        
        \addplot[fill=rendering,postaction={pattern=horizontal lines}]
                table[x expr = \coordindex, y = 4B,]{data/latency_normalized.dat}; 
        \addlegendentry{\em Big CPU (ARM-CL)}
        
        \addplot[fill=timedOutPointsColor,postaction={pattern=crosshatch}]
        table[x expr = \coordindex, y = 2S,]{data/latency_normalized.dat};
        \addlegendentry{\em Small CPU (ARM-CL)}
        
        \addplot[fill=cadmiumgreen,postaction={pattern=north west lines}]
        table[x expr = \coordindex, y = GPU,]{data/latency_normalized.dat};
        \addlegendentry{\em GPU (ARM-CL)}

        \addplot[fill=blizzardblue,postaction={pattern=grid},
        ]
        table[x expr = \coordindex, y = 4B+2S,]{data/latency_normalized.dat};
        \addlegendentry{\em Big+Small}

        \addplot[fill=unsatPointsColor,postaction={pattern=crosshatch dots},]
        table[x expr = \coordindex, y = 4B+GPU,]{data/latency_normalized.dat};
        \addlegendentry{\em Big+GPU}

        \addplot[fill=darkgoldenrod,postaction={pattern=north east lines},
        nodes near coords=
        {\pgfkeys{/pgf/fpu}\pgfmathparse{100-\pgfplotspointmeta*100}\pgfmathprintnumber{\pgfmathresult}},nodes near coords style = { /pgf/number format/.cd, fixed, fixed zerofill, precision=1, /tikz/.cd, font=\tiny\boldmath, color=darkgoldenrod, yshift=-0.1ex},
        ]
        table[x expr = \coordindex, y = 4B+2S+GPU,]{data/latency_normalized.dat};
        \addlegendentry{\em Big+Small+GPU (para-only)}

        \draw (1, 2.8) node[align=center] {\scriptsize \textcolor{darkgoldenrod}{Latency Imp. (\%): {\em para-only}}};

        \nextgroupplot[width=0.355\columnwidth, height=3.3cm,
                        ylabel={Normalized\\Throughput},
                        ylabel style={font=\scriptsize, align=center, text height=0pt},
                        xlabel style={font=\scriptsize, align=center, text height=0pt},
                        bar width=0.12,
                        ybar=0pt,    
                        ymin=0.25,
                        ymax=1.75,
                        ymajorgrids=true,
                        enlarge x limits={abs=0.55},
                        ytick={0,0.25,0.5,0.75,1,1.25,1.5},
                        yticklabels={0,0.25,0.5,0.75,1,1.25,1.5},
                        grid style=densely dashed,
                        typeset ticklabels with strut,]

        \addplot[fill=rendering,postaction={pattern=horizontal lines}]
                table[x expr = \coordindex, y = 4B,]{data/throughput_normalized.dat};       
        
        \addplot[fill=timedOutPointsColor,postaction={pattern=crosshatch}]
        table[x expr = \coordindex, y = 2S,]{data/throughput_normalized.dat};
        
        \addplot[fill=cadmiumgreen,postaction={pattern=north west lines}]
        table[x expr = \coordindex, y = GPU,]{data/throughput_normalized.dat};

        \addplot[fill=blizzardblue,postaction={pattern=grid},
        ]
        table[x expr = \coordindex, y = 4B+2S,]{data/throughput_normalized.dat};

        \addplot[fill=unsatPointsColor,postaction={pattern=crosshatch dots},]
        table[x expr = \coordindex, y = 4B+GPU,]{data/throughput_normalized.dat};

        \addplot[fill=darkgoldenrod,postaction={pattern=north east lines},nodes near coords=
        {\pgfkeys{/pgf/fpu}\pgfmathparse{\pgfplotspointmeta*100-100}\pgfmathprintnumber{\pgfmathresult}},nodes near coords style = { /pgf/number format/.cd, fixed, fixed zerofill, precision=1, /tikz/.cd, font=\tiny\boldmath, color=darkgoldenrod, yshift=-0.3ex, xshift=0ex},
        ]
        table[x expr = \coordindex, y = 4B+2S+GPU,]{data/throughput_normalized.dat};        

        \draw (1.1,1.6) node[align=center] {\scriptsize \textcolor{darkgoldenrod}{Throughput Imp. (\%): {\em para-only}}};

        \nextgroupplot[width=0.355\columnwidth, height=3.3cm,
                        ylabel={Normalized\\Energy Efficiency},
                        ylabel style={font=\scriptsize, align=center, text height=0pt},
                        xlabel style={font=\scriptsize, align=center, text height=0pt},
                        bar width=0.12,
                        ybar=0pt,    
                        ymin=0.5,
                        ymax=2.7,
                        ymajorgrids=true,
                        enlarge x limits={abs=0.55},
                        ytick={0,0.5,1,1.5,2,2.5},
                        yticklabels={0,0.5,1,1.5,2,2.5},
                        grid style=densely dashed,
                        typeset ticklabels with strut,]

        \addplot[fill=rendering,postaction={pattern=horizontal lines}]
        table[x expr = \coordindex, y = 4B,]{data/energy_efficiency_normalized.dat};       
        
        \addplot[fill=timedOutPointsColor,postaction={pattern=crosshatch}]
        table[x expr = \coordindex, y = 2S,]{data/energy_efficiency_normalized.dat};
        
        \addplot[fill=cadmiumgreen,postaction={pattern=north west lines}]
        table[x expr = \coordindex, y = GPU,]{data/energy_efficiency_normalized.dat};

        \addplot[fill=blizzardblue,postaction={pattern=grid}]
        table[x expr = \coordindex, y = 4B+2S,]{data/energy_efficiency_normalized.dat};

        \addplot[fill=unsatPointsColor,postaction={pattern=crosshatch dots},]
        table[x expr = \coordindex, y = 4B+GPU,]{data/energy_efficiency_normalized.dat};

        \addplot[fill=darkgoldenrod,postaction={pattern=north east lines},nodes near coords={\pgfkeys{/pgf/fpu}\pgfmathparse{\pgfplotspointmeta*100-100}\pgfmathprintnumber{\pgfmathresult}},nodes near coords style = { /pgf/number format/.cd, fixed, fixed zerofill, precision=1, /tikz/.cd, font=\tiny\boldmath, color=darkgoldenrod, yshift=-0.3ex},
        ]
        table[x expr = \coordindex, y = 4B+2S+GPU,]{data/energy_efficiency_normalized.dat};
        
        \draw (1.2,2.48) node[align=center] {\scriptsize \textcolor{darkgoldenrod}{Energy Effic. Imp. (\%): {\em para-only}}};

    \end{groupplot}           
\end{tikzpicture}
        \vspace*{-1mm}
        \caption{Normalized latency, throughput, and energy efficiency for coarse-grained non-pipelined, parallel co-execution on Amlogic SoC relative to sequential execution on its big CPU cluster.}
        \label{fig:co_executing_units}
    \end{minipage}
\end{figure*}
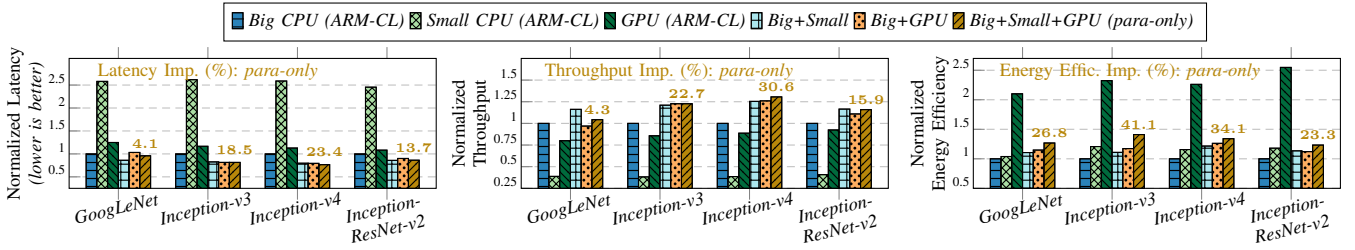

\begin{figure}[tb]
    \raggedleft
    \begin{tikzpicture}
    \centering
    \pgfplotstableread{
        name
        {\em GoogLeNet}
        {\em Inception-v3}
        {\em Inception-v4}
        {\em Inception-ResNet-v2}
        {\em GoogLeNet}
        {\em Inception-v3}
        {\em Inception-v4}
        {\em Inception-ResNet-v2}
        {\em GoogLeNet}
        {\em Inception-v3}
        {\em Inception-v4}
        {\em Inception-ResNet-v2}
        }\datatable
        
    \begin{groupplot}[
        group style={
            group name=my plots,
            group size=1 by 1,
            xlabels at=edge bottom,
            xticklabels at=edge bottom,
        },
        ytick=data,
        yticklabels from table={\datatable}{name},
        y tick label style={anchor=east,font=\tiny,rotate=0,text height=0pt,text width=1.5cm,align=center,,inner sep=0pt,outer sep=1pt},
        xlabel={Latency (s)},
        x tick label style={font=\tiny,rotate=0,text height=0.1pt,inner sep=3pt,outer sep=2pt},
        ylabel style={font=\scriptsize,rotate=270,align=center,text height=0.1pt,inner sep=0pt,outer sep=0pt},
        xlabel style={font=\scriptsize, align=center, text height=0.1pt,inner sep=0pt,outer sep=1.5pt},]

        \nextgroupplot[width=0.8\columnwidth,, height=4.1cm,
                        xbar stacked,
                        bar width=3.5pt,
                        xmin=30,
                        xmax=800,
                        xmajorgrids=true,
                        grid style=densely dashed,
                        xtick={0,100,200,300,400,500,600,700,800},
                        xticklabels={0,0.1,0.2,0.3,0.4,0.5,0.6,0.7,0.8},
                        typeset ticklabels with strut,
                        legend style={legend columns=3, at={(0.5,1.08,1)}, anchor=south, font=\scriptsize},
                        legend image code/.code={%
                           \draw[yshift=-0.25em]
                            (0cm,0cm) rectangle (4pt, 0.6em);},]

                    \addplot[fill=blizzardblue,y tick label style={yshift=-0.1cm},postaction={pattern=vertical lines}] table[y=Clusters,x=comp] {data/finegrained_overhead.dat};
                    \addlegendentry{\em Comp.};

                    
                    \addplot[fill=unsatPointsColor,y tick label style={yshift=-0.1cm},postaction={pattern=north east lines}
                    ] table[y=Clusters,x=sync] {data/finegrained_overhead.dat};
                    \addlegendentry{\em Sync. Overhead};
                    
                    \addplot[fill=timedOutPointsColor,y tick label style={yshift=-0.1cm},postaction={pattern=crosshatch}] table[y=Clusters,x=code] {data/finegrained_overhead.dat};
                    \addlegendentry{\em Program Overhead};  

                    \draw (650,14.5) node[align=center] {\scriptsize \bfseries \textcolor{cadmiumgreen}{{\em 11.7\%}}};

                    \draw (670,8.5) node[align=center] {\scriptsize \bfseries \textcolor{deeperorange}{{\em 16.5\%}}};

                    \draw (510,7.5) node[align=center] {\scriptsize \bfseries \textcolor{deeperorange}{{\em 21.6\%}}};

                    \draw (250,6.5) node[align=center] {\scriptsize \bfseries \textcolor{deeperorange}{{\em 19.5\%}}};

                    \draw (100,5.5) node[align=center] {\scriptsize \bfseries \textcolor{deeperorange}{{\em 29.2\%}}};

    \end{groupplot} 
    \draw (-2.2,2)  node{\scriptsize{\em Big+Small+GPU}};
    \draw (-2.15,1.2)  node{\scriptsize{\em Big+GPU}};
    \draw (-2.15,0.5)  node{\scriptsize{\em Big+Small}};
 
\end{tikzpicture}
    \vspace*{-4mm}
    \caption{Latency distribution of fine-grained mapping for non-pipelined, parallel co-execution on Amlogic SoC: computation, synchronization overhead, and program overhead.}
    \label{fig:impl_overhead}
\end{figure}
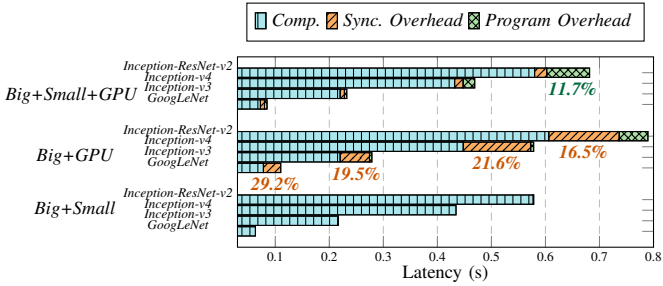

\subsection{Co-execution of Heterogeneous Units}
\label{subsec:co-diverse-units}
This section evaluates the performance of coarse- and fine-grained mapping algorithms in non-pipelined parallel execution via various on-chip processor combinations to demonstrate the efficacy of the operator mapping approaches.
We implement and test the strategies generated to ensure a robust analysis, collecting corresponding experimental data on an Amlogic SoC, specifically using a real Khadas Vim3 Pro platform.

\subsubsection{Coarse-grained Mapping}
Fig.~\ref{fig:co_executing_units} illustrates performance and energy gains achieved through coarse-grained co-execution mapping strategies: CPU clusters, big CPU cluster with GPU, and CPU clusters with GPU, relative to the highest-performance big CPU cluster. Parallel processing with CPU clusters obtains an average latency decrease of approximately 16.4\% and a throughput increase of about 19.8\%. However, the setup with a big CPU and a GPU only marginally outperforms the dual CPU cluster configuration, even if the small CPU cannot deliver even half the performance of the GPU. For models exhibiting lower degrees of parallelism, such as GoogLeNet and Inception-ResNet-v2, the advantages of CPU-GPU co-execution were offset by communication overheads, making it less beneficial. Despite utilizing two CPU clusters and a GPU for parallel inference, the improvement was relatively limited compared to the dual CPU cluster option. Thus, employing purely operator-based parallel execution using heterogeneous on-chip computing resources is not advisable except where extreme latency reduction is required. 

Regarding energy efficiency, since the small CPU and GPU are specifically optimized for power efficiency, integrating a big CPU with either a small CPU, GPU, or both can significantly boost energy efficiency. The corresponding combinations deliver average energy efficiency gains of 14.2\%, 17.6\%, and 31.3\%, respectively. These improvements primarily stem from reduced reliance on the power-intensive big CPU cluster and reduced execution times by engaging more processors. However, while the GPU is recognized for its exceptional power efficiency, its combined use with CPUs somewhat dilutes these benefits.

Consequently, the pipe-only and various hybrid configurations in {\em Para-Pipe}, which integrate sequential processing stages featuring the GPU, more effectively preserve the GPU’s contribution to power efficiency. Besides, selecting processors with compatible computing architectures and data formats, like big and small CPU clusters, for parallel processing within stages is more advantageous. These strategies enhance throughput with minimal latency trade-offs and reduce unnecessary communication and data conversion overhead, thus improving the efficient and effective utilization of on-chip resources.

\subsubsection{Fine-grained Mapping} 
\paragraph{Overhead Analysis}
We assess latency distributions for fine-grained mapping strategies by comparing them to those obtained from coarse-grained mappings across both runtimes. 
Program overhead, defined as the additional runtime introduced by the implementation complexity of fine-grained mapping, is estimated as the difference between coarse-grained mapping performance on coarse- and fine-grained runtimes. The computational cost associated with fine-grained mapping is reflected in the lower execution times observed for both strategies. Synchronization overhead is calculated as the difference between the execution time of a fine-grained strategy and the sum of its program overhead and computational cost.

As illustrated in Fig.~\ref{fig:impl_overhead}, fine-grained mapping incurs an average program overhead of 5.9\% during the concurrent execution of CPU clusters and GPUs, notably impacting large-scale model inferences, such as the 11.7\% overhead observed in Inception-ResNet-v2. 
Synchronization overhead averages 21.7\% across tested models when using a big CPU cluster and GPU but decreases significantly to 4.7\% with the addition of a small CPU cluster. The inclusion of the small CPU cluster redistributes operators more evenly, alleviating GPU workloads and reducing CPU-GPU synchronization points by 20.1\% on average. This, in turn, minimizes CPU-GPU communication and streamlines coordination. Conversely, configurations with only a big CPU cluster and GPU experience frequent and complex CPU-GPU communications. These are further exacerbated by the asynchronous and intermittent command submissions to the \textit{OpenCL} queue, resulting in execution jitters that disrupt synchronization efficiency.

Despite these challenges, fine-grained mapping is markedly more effective for aligning irregular model structures with units that share similar architectures and data formats, as demonstrated with two CPU clusters in our study. In scenarios involving the co-execution of two CPU clusters, the fine-grained approach circumvents the complexities of synchronization and program implementation, yielding an average improvement of 3.35\% in latency and 4.28\% in throughput compared to coarse-grained mapping.

\paragraph{Operator Mapping Choice}
The coarse-grained method rapidly generates parallel operator mapping strategies for graphs with independent branches, minimizing processor switching and complex synchronization requirements. This approach is remarkably robust against execution jitters and runtime interruptions from the operating system, making it well-suited for more volatile environments. Conversely, when applied to these relatively straightforward graphs, the fine-grained method performs best in stable runtime environments with uniform computing architectures. Moreover, it excels in mapping complex and irregular model graphs with solution times that scale appropriately with the degrees of operator parallelism.


\subsection{Mapping time Analysis}
The dual parallel mapping methods of {\em Para-Pipe} enhance the parallel scheduling of subgraphs by generating multiple ILP problems, thereby increasing the efficiency of scheduling large models. 
Due to its detailed granularity, fine-grained mapping inherently demands longer resolution times than coarse-grained mapping. Consequently, this discussion concentrates on fine-grained mapping to comprehensively assess the scheduling time overhead associated with our framework.

Table~\ref{tab:ilp_time} presents solution times for the fine-grained algorithm when generating the para-only option on the BST SoC. Once parallel operator mapping strategies are established for each subgraph, no further ILP computations are necessary. {\em Para-Pipe} utilizes these results to efficiently generate additional mapping options, such as hybrid-T and hybrid-L, with minimal overhead. Thus, ILP times shown in Table~\ref{tab:ilp_time} approximate the total scheduling time required to generate all Pareto-optimal points.

By constructing subgraph-based ILP problems, we can rapidly devise efficient mapping solutions for most networks within 5 minutes. For the PETR-based network, the largest subgraph contains 169 operators and exhibits an inter-operator parallelism degree of 16, making it the most complex subgraph analyzed in this study (its partial architecture with 67 operators is detailed in Fig.~\ref{fig:model-structure-demo}). The ILP solution time for this subgraph is approximately 6 hours. For large-scale models with multiple complex subgraphs, such as those found in PETR-based networks, subgraph mappings can be scheduled in parallel to expedite the solution process significantly. Unlike meta-heuristic or heuristic algorithms, our subgraph-based ILP approach swiftly generates optimal results for each subgraph without requiring time-consuming iterative adjustments. 
This approach ensures the optimal solution is found within a tolerable timeframe, even for highly intricate subgraphs.

\setlength\tabcolsep{3pt}
\begin{table}[tb]
    \centering
    \caption{ILP solution time (minutes) for fine-grained mapping algorithm for non-pipelined, parallel execution with NPU and two DSPs (\textit{para-only}) on BST SoC: Performance Evaluation on Intel(R) Xeon(R) Gold 6326 CPU @ 2.90GHz}
    \label{tab:ilp_time}
        \begin{tabular}{|l|c|c|r|}
        \hline
        \multicolumn{1}{|c|}{Network}&\makecell{\#Subgraphs}&\makecell{\#Operators}&\makecell{Time (minutes)}\\
            \hline
            \hline
            GoogLeNet& 11 & 141 & $<1$\\
            \hline
            Inception-v3& 13 & 220 & 4\\
            \hline
            Inception-v4& 21 & 343 & 4\\
            \hline
            Inception-ResNet-v2& 45 & 576 & $<$1\\
            \hline
            PETR-based& 24 & 337 & 361\\
            \hline
            BEVFormer-based& 25 & 257 & $<$1\\
        \hline
        \end{tabular}
\end{table}
\section{Related Work}
Optimizing ML inference on heterogeneous SoCs has garnered extensive research interest. Brakel et al.~\cite{brakel2024model} have explored model parallelism within distributed infrastructures, focusing on intra- and inter-operator parallelism, and highlighted their findings through case studies on multi-billion parameter transformer models.
This paper aims to explore ML inference on SoCs further, concentrating on the following key dimensions:

\paragraph{Pipelining}
\textit{Pipe-it}\cite{wang2019high} exploits inter-operator parallelism through pipelining to engage \textit{ARM} \textit{big} and \textit{LITTLE} CPU clusters concurrently, enhancing throughput. Similarly, PipeBERT\cite{chang2023pipebert} focuses on pipelining BERT models onto heterogeneous CPU clusters, employing an improved binary search algorithm to optimize pipeline configurations. Mukherjee et al.\cite{mukherjee2022automated} and Flexi-BOPI\cite{wang2024flexi} utilize dynamic programming and Bayesian optimization respectively to automate the partitioning process. 
While these strategies significantly enhance throughput, they do not always ensure acceptable execution latency, especially when applied to large-scale modern neural networks. This limitation poses challenges for applications where low latency is critical.

\paragraph{Operator Parallel Execution} 
Tumeo et al.~\cite{tumeo2008ant} introduce an \textit{Ant Colony Optimization} algorithm for mapping tasks onto heterogeneous multiprocessor architectures. A Mixed ILP model is developed in~\cite{bender1996milp} to optimize task mapping by considering the data exchange overhead among processors. 
Topcuoglu et al.~\cite{topcuoglu2002performance} propose two classic DAG mapping algorithms, utilizing defined upward and downward rank values to prioritize task scheduling. Zhang et al.~\cite{zhang2015maximizing} devise reliable scheduling for parallel task execution in heterogeneous computing systems, emphasizing both task reliability and energy efficiency.
These methodologies are primarily designed for handling task graphs with limited operator parallelism and assume the presence of random and irregular structures, which do not always align with the structured nature of many real-world applications. Additionally, DUET~\cite{zhang2021duet} partitions complex-structured DNNs into subgraphs and employs a greedy-correction subgraph scheduling algorithm for near-optimal placement across coupled CPU-GPU architectures. This approach primarily targets multi-model tasks without delving into fine-grained operator-level parallelism. 

\paragraph{Intra- and Inter-Operator Parallelism}
Minakova et al.~\cite{minakova2020combining} generate the execution pipeline using CPUs at the operator-level granularity, with GPUs accelerating heavy computation within operators. Similarly, Jeong et al.~\cite{jeong2021deep} highlight parallelization using GPU and NPU via pipelining and multithreading on TensorRT. \textit{Synergy}~\cite{zhong2019synergy} proposes a dynamic runtime strategy, work-stealing, to adjust workload mapping configuration on SoCs. Despite these advancements, the focus remains predominantly on linear models, leaving room for optimization in more complex and irregular model architectures.

\paragraph{Multiple Workloads}
Kim et al.~\cite{kim2022energy} present an energy-aware pipeline-based mapping methodology for multiple DL applications onto multi-processors. They focus on the dynamic scheduling of applications to reduce energy consumption via DVFS under latency constraints, with each application being a simple and linear graph. 
Moreover, DAIMO-NPU~\cite{sohn2024strategy} adopts a cyclic scheduling algorithm to support dynamic addition and removal of model execution tasks onto heterogeneous NPU settings. BAND~\cite{jeong2022band} orchestrates multi-DNN workloads on heterogeneous compute engines, partitioning models into subgraphs and dynamically scheduling these subgraphs for optimized execution. These strategies are crafted for scenarios involving the scheduling of multiple applications under latency constraints without extensively exploring the intricate model structures characteristic of single, complex, and computation-intensive application inferences.

Our work innovatively addresses the processing of irregular modern model graphs by leveraging hierarchical operator parallelism within a pipelined architecture. This meticulous tuning of inter- and intra-stage operator parallelism significantly broadens the scope of ML inference optimization on heterogeneous SoCs. Our method achieves a balanced trade-off between latency and throughput and boosts the platform's energy efficiency. Additionally, it generates multiple Pareto-optimal strategies, accommodating diverse user preferences across various performance metrics.
Table~\ref{tab:compare_related} underscores the critical distinctions between our methodology and existing works, highlighting our contributions to the field.

\setlength\tabcolsep{2.5pt}
\begin{table}[tb]
    \centering
    \caption{Comparing \textit{Para-Pipe} with other frameworks that support ML inference on heterogeneous SoCs}
    \label{tab:compare_related}
    \resizebox{\linewidth}{!}{
        \begin{tabular}{|c|c|c|c|c|c|}
        \hline
        \multicolumn{1}{|c|}{Framework}&\makecell{Irregular\\Model Graph$^{\mathrm{a}}$}&\makecell{Pipelining}&\makecell{Operator Parallel\\Execution}&\makecell{Strategy\\Selection$^{\mathrm{b}}$}\\
            \hline
            \hline
            ~\cite{zhang2021duet,tumeo2008ant,bender1996milp,topcuoglu2002performance,zhang2015maximizing}& \xmark & \xmark & \cmark & \xmark\\
            \hline
            ~\cite{wang2019high,wu2020pipeline,minakova2020combining,chang2023pipebert,mukherjee2022automated,aghapour2023pipelined,wang2024flexi,aghapour2024arm}& \xmark & \cmark & \xmark & \xmark\\
            \hline
            \textbf{\textit{Para-Pipe}}& \cmark & \cmark & \cmark & \cmark\\
        \hline
        \end{tabular}
    }
        \\
        {\scriptsize
        $^{\mathrm{a}}$A complex neural network with intensive operator parallelism and interdependencies, unlike simpler linear networks like MobileNet and ResNet50.
        $^{\mathrm{b}}$Selection of mapping strategies based on prioritized metrics like latency or energy efficiency.}
\end{table}



\section{Conclusion}
This paper presents {\em Para-Pipe}, a novel partitioning and scheduling framework that optimizes the trade-offs between latency and throughput to maximize energy efficiency on embedded computing platforms. {\em Para-Pipe} efficiently maps modern neural networks across heterogeneous processors by leveraging hierarchical operator parallelism within computational graphs.
The proposed methodology decomposes the model graph into subgraphs with varying degrees of operator parallelism and assigns them to different computing units for parallel or sequential execution in distinct pipeline stages.
This adaptive fusion of parallel and sequential execution enables fine-grained control over latency-throughput trade-offs, enhancing overall energy efficiency.
Experimental evaluations on two distinct SoCs (Amlogic and BST) demonstrate {\em Para-Pipe}'s ability to generate Pareto-optimal configurations, balancing latency and throughput while identifying the most energy-efficient mapping strategies for networks with dense inter-operator parallelism.

\bibliographystyle{IEEEtran}
\bibliography{refs}

\end{document}